\documentclass[a4paper,11pt]{article}
\pdfoutput=1 
\usepackage{jcappub} 
\usepackage[T1]{fontenc}
\usepackage{appendix}
\usepackage[dvipsnames]{xcolor}
\usepackage{bm}
\usepackage{comment}
\usepackage{chngcntr}
\usepackage{soul}

\def\msun{h^{-1}{M_\odot}}
\def\mpc{h^{-1}{\rm Mpc}}
\def\kpc{h^{-1}{\rm kpc}}
\def\hmpci{h{\rm Mpc}^{-1}}
\def\bx{\bm{x}}

\title{Power Spectrum of Intrinsic Alignments of Galaxies in IllustrisTNG}

\author[a,1]{Jingjing~Shi,\note{Corresponding author.}}
\author[a,b]{Toshiki~Kurita,}
\author[a]{Masahiro~Takada,}
\author[c]{Ken~Osato,}
\author[a,b]{Yosuke~Kobayashi,}
\author[d,a]{Takahiro~Nishimichi}

\affiliation[a]{Kavli Institute for the Physics and Mathematics of the Universe (WPI), The University of Tokyo Institutes for Advanced Study (UTIAS),
The University of Tokyo, 5-1-5 Kashiwanoha, Kashiwa-shi, Chiba, 277-8583, Japan}
\affiliation[b]{Department of Physics, The University of Tokyo, 7-3-1 Hongo, Bunkyo-ku, Tokyo 113-0033 Japan}
\affiliation[c]{Institut d'Astrophysique de Paris, Sorbonne Universit\'{e}, CNRS, UMR 7095, 75014 Paris, France}
\affiliation[d]{Center for Gravitational Physics, Yukawa Institute for Theoretical Physics, Kyoto University,
Kyoto 606-8502, Japan}

\emailAdd{jingjing.shi@ipmu.jp}

\abstract{We present the 3-{\it dimensional} intrinsic alignment power spectra between the projected 2d galaxy shape/spin and the 3d tidal field
across $0.1<k/\hmpci<60$ using cosmological hydrodynamical simulation, Illustris-TNG300, at redshifts ranging from $0.3$ to $2$. The shape-tidal field alignment increases with galaxy mass and the linear alignment coefficient $A_{\rm IA}$, defined with respect to the primordial tidal field, is found to have weak redshift dependence. We also show a promising detection of the shape/spin-tidal field alignments for stellar mass limited samples and a weak or almost null signal for star-forming galaxies for the TNG300 volume, $\sim 0.01~(h^{-1}{\rm Gpc})^3$.
We further study the morphology and environmental dependence of the intrinsic alignment power spectra. The shape of massive disk- and spheroid-galaxies tend to align with the tidal field. The spin of low mass disks (and spheroids at low redshifts) tend to be parallel with the tidal field, while the spin of massive spheroids and disks tend to be perpendicular to tidal field. The shape and spin of massive centrals align with the tidal field at both small and large scales. Satellites show a radial alignment within the one-halo term region, and low mass satellites have an intriguing alignment signal in the two-halo term region. We also forecast a feasibility to measure the intrinsic alignment power spectrum for spectroscopic and imaging surveys such as Subaru HSC/PFS and DESI. 
Our results thus suggest that galaxy intrinsic alignment can be used as a promising tool for constraining the galaxy formation models.}

\begin{document}
\maketitle
\flushbottom

\section{Introduction}
\label{sec:intro}
The shape of galaxies is modulated by the surrounding large scale structure primarily through gravitational interaction. Therefore, galaxies tend to align with the large scale structure and the other galaxies that lie in the same large scale structure, a phenomenon known as intrinsic alignment (IA). With the advent of cosmological weak lensing survey, IA has been extensively studied in the literature, mainly motivated by the fact that it is one of the dominant systematic effects that affect cosmology inference from weak lensing measurements
\cite{2007NJPh....9..444B,2007A&A...464..399H,2010A&A...523A...1J,2011MNRAS.410..844M,2011A&A...527A..26J,2016MNRAS.456..207K,2012JCAP...05..041B} {(see Ref.~\cite{2015SSRv..193...67K,TroxelandIshak2015} for a review). Future weak lensing cosmology requires a better understanding of the IA effects} for galaxies of various types from large scales down to small scales into one-halo term region. However, our knowledge on IA, especially on small scales and for high redshift galaxies, are still limited.
Rather than being a contamination, IA itself can be a useful probe because it carries cosmological information \cite{2013JCAP...12..029C,2020arXiv200703670A} and it can be used to constrain the galaxy formation and evolution processes \cite{Kiesslingetal2015,Joachimietal2015}. 

The IA can be induced by either the stretching of the system along the tidal field \cite{2001ApJ...559..552C,HirataSeljak2004}, or tidal torquing that spins the system up \cite{1969ApJ...155..393P, 1984ApJ...286...38W}. Theoretical {modeling} of galaxy IA is developed based on these physical pictures, including linear alignment model \cite{HirataSeljak2004}(hereafter LA), non-linear alignment model \cite{2007NJPh....9..444B} (hereafter NLA), and quadratic alignment model \cite{2001ApJ...559..552C}(hereafter QA). It is usually assumed that LA and NLA models apply to elliptical galaxies, while the QA model applies to disk galaxies. However, the above models mainly work at linear scales (the NLA model works at mildly non-linear scales by a phenomenological extension of LA model), the small scale IA is complicated by non-linearity, baryonic physics, and the inclusion of satellites. Besides, Ref.~\cite{2020JCAP...01..025V} gives an effective field theory description for IA.
Further investigations are clearly needed for constraining the current IA models and further development.

Various works have studied 
either the alignment of galaxy shapes or the alignment of angular momentum. For example, the shape alignment is found to be stronger for massive/luminous galaxies \cite{2019MNRAS.485.2492C}, with no significant redshift evolution \cite{2013MNRAS.431..477J,2015MNRAS.450.2195S}.
Other studies focused on the color, morphology, or environmental dependence of IA. Luminous red galaxies show a strong shape-alignment \cite{2011A&A...527A..26J}, while blue galaxies show no shape-alignment detection \cite{2011MNRAS.410..844M}. It was shown that spheroidal galaxies tend to lie along the filament \cite{2016MNRAS.462.2668T,2019MNRAS.488.5580P}, with their spin preferentially being perpendicular to the filament \cite{2018MNRAS.481.4753C}. One the other hand, Ref.~\cite{2016MNRAS.462.2668T} showed a null detection of IA for disk galaxies using the two-point correlation function, while other works \cite{2011ApJ...732...99L,2013ApJ...775L..42T} indicated that the spin of disks appears to be aligned with the large scale structure.
The environmental dependence of shape/spin alignment has also been studied for subsample of galaxies divided based on their environments such as centrals and satellites \cite{2015MNRAS.448.3522T,2015MNRAS.450.2195S,2018MNRAS.481.4753C,2013MNRAS.431..477J}. Although no consensus has been reached on the satellite alignment at large scales, a radial alignment of satellites at small scales, i.e., satellites tend to orientate towards the centrals, is confirmed in both simulation and observations.

As discussed above, for different galaxy samples, different mechanism dominates their IA signal. Measuring IA using only the shape inferred from the mass distribution or angular momentum might miss the signal induced by the other one in the galaxy samples. We will present here, a measurement of IA for the same galaxy samples, using both shape and angular momentum of individual galaxies in a realistic numerical simulation.

Cosmological hydrodynamical simulations provide a natural way to study the complex interplay between galaxy shape, angular momentum, and large scale structures. Various physical processes that affect the galaxy shape/orientation, including the large scale tidal field, gas accretion, mergers, and feedback, could be all included self-consistently in hydrodynamical simulations. Such simulations covering a cosmological volume also enable us to study statistically the IA effect for galaxies divided based on their properties such as stellar mass, morphology, and environments. Comparison of the simulation results with observations can provide extra constraints on the input physics in the simulation.  

Most recently, Ref.~\cite{2020arXiv200105962T} developed a new method that allows us to measure the IA power spectrum in {\it three-dimensional} (3D) space to quantify the IA effects of halo shapes measured from cosmological $N$-body simulations. The scale dependence is naturally included in the IA power spectrum. Compared to the widely used projected two-point correlation function, the IA power spectrum in 3D space has the merit of maintaining the full information at the level of two-point statistics. 
There are promising prospects in the future; by combining imaging data such as Subaru HSC \citep{2018PASJ...70S...4A} and DES \footnote{\url{https://www.darkenergysurvey.org}}
with the existing and upcoming spectroscopic surveys such as BOSS \cite{2013AJ....145...10D}, eBOSS \cite{2016AJ....151...44D},
DESI \cite{2016arXiv161100036D} and Subaru PFS \cite{2014PASJ...66R...1T} for the overlapping regions of their survey footprints, we should be able to measure the 3D IA power spectrum.
Hence the purpose of this paper is to make a quantitative study of the IA effects for various types of galaxies using the state-of-the art hydrodynamical simulations, Illustris-TNG300 \cite{2018MNRAS.475..676S,2019ComAC...6....2N}. Our study will give us a guideline for measurements of the IA effects based on the 3D power spectrum method, for the existing and upcoming datasets. To do this, we study the IA effects of galaxies over the range of scales, $0.1< k/\hmpci <60$ for galaxies at $0.3\le z\le 2$.

In this work, we will study how the IA effects vary with different types of galaxies (massive/star-forming galaxies),  galaxy morphology (spheroids/disks) , and the environments (central/satellite galaxies). We will also study how the IA effects change with wavenumber ($k$) and redshifts. Our results would be useful to constrain the current models of galaxy IA effects.
Our study also gives guidance on how to use the dependencies of the measured IA effects on galaxy properties to explore galaxy physics and cosmological constraints.

The structure of this paper is as follows.
The simulation and galaxy selection is described in Section~\ref{sec:method}. In Section~\ref{sec:formalism}, we briefly introduce the non-linear alignment model and the quadratic alignment model. Galaxy IA for different stellar mass and its evolution across the redshift of $0.3$ to $2$ are presented in Section~\ref{subsec:mstar_z}. In Section~\ref{subsec:IA_fixng}, we study the IA power spectrum for $n_g=10^{-4} \, (\mpc)^{-3}$ galaxy samples ranked either by $M_\star$ or SFR. In Sections~\ref{subsec:IA_morph} and \ref{subsec:IA_censat}, we explore the dependence of IA on galaxy morphology and environment (central/satellite). We further present a prediction of IA for future surveys, including their signal-to-noise ratio in Section \ref{sec:survey}.

\section{Illustris-TNG and Methods}
\label{sec:method} 

\begin{figure}[!htb]
\centering 
\includegraphics[width=.95\textwidth]{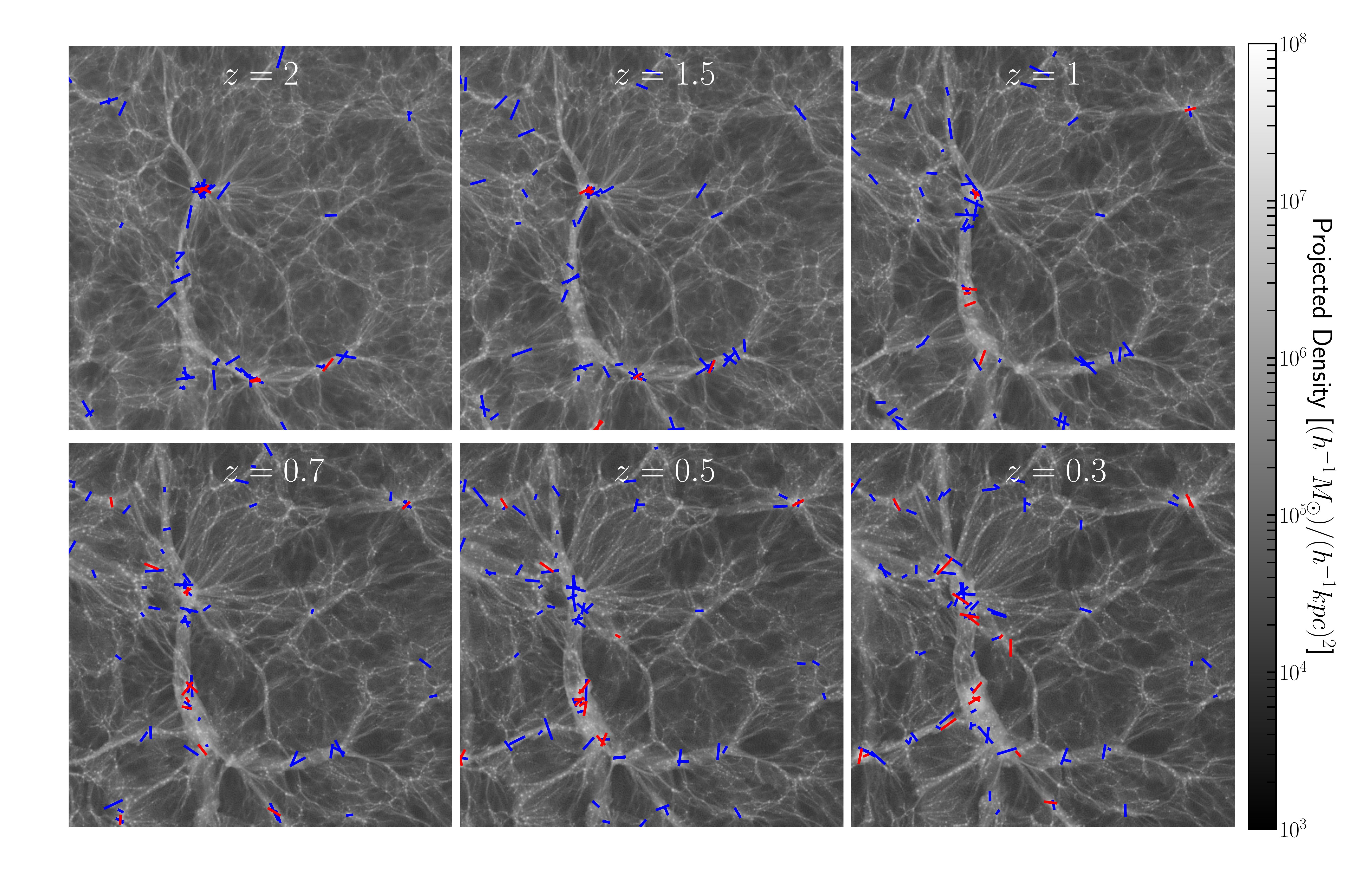}
\caption{Snapshots of a $40\times 40(\mpc)^2$ slice with a thickness of $2\mpc$ at $z=2$, 1.5, 1, 0.7, 0.5, and 0.3, respectively. The blue and red sticks represent star-forming and quiescent galaxies ($M_\star>10^9\msun$), where the quiescent galaxies are selected by applying ${\rm sSFR}<10^{-11}M_\odot~{\rm yr}^{-1}$. The direction of each stick is defined to be along the major axes of the projected ellipse of the galaxy and the length is proportional to the ellipticity amplitude.}
\label{fig:snapshot}
\end{figure}
\subsection{The {\rm Illustris-TNG} simulations}
\label{subsec:sim}

Throughout this work, we use the data from the Illustris-TNG project \cite{2019MNRAS.490.3234N}.
Illustris-TNG is a suite of cosmological magneto-hydrodynamical (MHD) simulations that feature comprehensive models for galaxy formation physics, including primordial and metal-line radiative cooling in the presence of a time dependent background ionization field, star formation, ISM physics \cite{2003MNRAS.339..312S},
stellar evolution and the chemical enrichment\cite{2018MNRAS.477.1206N}, supernova feedback, black hole growth and the feedback that happens along the black hole accretion \cite{2017MNRAS.465.3291W}, and also magnetic field under the assumption of ideal MHD \cite{2013MNRAS.432..176P}
\citep[for more details on the TNG model, see ][]{2018MNRAS.479.4056W,2018MNRAS.473.4077P,2018MNRAS.480.5113M,2020arXiv200509654N}.
The TNG simulations are performed using the moving-mesh code {\tt AREPO} \citep{2010ARA&A..48..391S}, which follows the evolution of dark matter, baryons, and black holes from $z=127$ to $z=0$. The TNG project includes simulations of three different volumes with the same initial conditions and physical models: TNG50, TNG100, and TNG300, where the side length of the boxes are approximately 
$50$~Mpc, $100$~Mpc, and $300$~Mpc (without $h$ in units), respectively. In this work, we use the largest volume run, TNG300, that includes $2\times 2500^3$ resolution elements, giving an average gas cell mass of $7.44 \times 10^6\msun$, a dark matter mass resolution of $3.98 \times 10^7\msun$, and a collision-less softening length of $1\kpc$. The simulations adopt the {\it Planck} cosmology \citep{planck-collaboration:2015fj}: $\Omega_{\rm m}=0.3089$, $\Omega_{\rm b}=0.0486$, $\Omega_{\rm \Lambda}=0.6911$, $\sigma_8=0.8159$, $n_s=0.9667$, and $h=0.6774$. 

The simulations fairly well reproduce various observational results. For example, they are calibrated to reproduce the cosmic star formation rate density at $z\leq10$, the observed galaxy stellar mass function, the stellar-to-halo mass relation, the total gas mass content within the massive groups, the stellar mass--stellar size relation, and the BH--galaxy mass relations at $z=0$ \citep{2018MNRAS.473.4077P}.
Besides those used to calibrate the models, the predictions of simulations match a range of observations, such as the galaxy stellar mass functions up to $z\sim 4$ \citep{2018MNRAS.475..648P},
the galaxy clustering of blue and red galaxies \citep{2018MNRAS.475..676S}, stellar sizes up to $z\sim2$ 
\citep{2018MNRAS.474.3976G}, etc.

Dark matter halos and subhalos and galaxies in each parent halo are identified by using the Friends-of-Friends (FoF) and \textsc{subfind} algorithms \citep{1985ApJ...292..371D,2001NewA....6...79S}. Usually, each FoF group contains one or more \textsc{subfind} structures (hereafter substructure or subhalo) and the baryonic component in a substructure is defined as a galaxy. The most massive substructure within a FoF group is usually refereed as the central galaxy, while the others (if exist) are satellites.

\subsection{Galaxy sample selection}
\label{subsec:gal_sample}

We aim to study the galaxy IA across a variety of galaxy properties over redshifts ranging from $z=0$ to $z=2$. We select all galaxies with stellar mass $M_\star>10^9\msun$ at $z=0.3$, $0.5$, $0.7$, $1$, $1.5$, and $2$, but exclude those with {\it SubhaloFlag=0} \cite{2019ComAC...6....2N}
which are possibly a baryonic fragment of a disk or other galactic structure identified by \textsc{subfind} yet formed within a halo.
In this work, the stellar mass of a galaxy is the summing up of all gravitationally bound stellar particles within twice the half stellar mass radius $R_{\star,1/2}$. To study the dependence of IA on galaxy mass, we divide the sample into three stellar mass ranges: $M_\star/\msun{\in}[10^9,10^{10}], [10^{10}, 10^{11}]$ and $[10^{11},10^{12}]$, respectively.
To understand how the IA signal varies with environment and galaxy morphology, we study centrals/satellites and disk/spheroid galaxies separately. We use $\kappa_{\rm rot}$, which is defined as the ratio of the rotational kinetic energy to the total kinetic energy 
\cite{2012MNRAS.423.1544S,2017MNRAS.467.3083R}, 
$\kappa_{\rm rot}=\frac{K_{\rm rot}}{K}=\frac{1}{K}\sum_i\frac{1}{2}m_i v^2_{i, {\rm rot}}$, to distinguish the galaxies of different morphology types, where $v_{i,{\rm rot}}$ is the rotational velocity of stellar particle with respect to the galaxy center. We choose galaxies with $\kappa_{\rm rot}>0.55$ as disk-like, rotation dominated galaxies, and galaxies with $\kappa_{\rm rot}<0.45$ as spheroid-like, dispersion dominated galaxies. Tables~\ref{tab:sample_mstar}--\ref{tab:sample_central} summarizes the galaxy samples used in this work.
Note that the number of galaxies in each sample can be obtained by multiplying the number density, given in the table, and the volume of the simulation ($V=(302~{\rm Mpc})^3\simeq (205~h^{-1}{\rm Mpc})^3)$. 
For instance, we have $393$ galaxies for the stellar mass limited sample with $10^{11}<M_\star/\msun<10^{12}$ at $z=2$. 

Fig.~\ref{fig:snapshot} shows the snapshots of a $40\times 40 \times 2 (\mpc)^3$ comoving volume from redshift $2$ to $0.3$. The blue and red sticks indicate the star-forming and quiescent galaxies, respectively. The major axis of the projected ellipticity of each galaxy is along the stick, where the length of 
the stick is defined to be proportional to the ellipticity amplitude. 
At $z=2$, most of galaxies are star-forming galaxies. These star-forming galaxies tend to reside in the filaments, while the quiescent galaxies live in the intersecting knots of filaments.
As large-scale structures evolve according to the gravitational instability at lower redshifts, more and more galaxies are lying in the knots of the cosmic web and turn to quiescent.

To make a comparison with ongoing/future surveys, such as PFS and DESI, we apply a cut in either stellar mass or star formation rate (SFR). 
We use the SFR as a proxy of emission-line strength because emission-line galaxies (ELGs) are one of the targets in the future surveys. A cut in SFR in the total model galaxy population roughly corresponds to the selection of ELGs based on [OII] emission lines strength \cite{2020arXiv200106560G} (also see Osato et al. in preparation).
Here SFR of each galaxy in the simulation is defined by the spontaneous SFR summed up within the whole galaxy. In Fig.~\ref{fig:ng_fixed} we explicitly show the cuts to define the galaxy samples we use in this paper.

\begin{table}
\centering
\scriptsize
\begin{tabular}{l|p{2.5em}p{1.5em}p{2.5em}p{5em}|p{2.5em}p{1.5em}p{2.5em}p{5em}|p{2.5em}p{1.5em}p{2.5em}p{5em}} \hline\hline 
 & \multicolumn{4}{c|}{$M_\star{\in}[10^9,10^{10}]$} & \multicolumn{4}{c|}{${M_\star \in}[10^{10},10^{11}]$}& \multicolumn{4}{c}{${M_\star \in}[10^{11},10^{12}]$} \\
 \hline
 $z$& $10^2n_g$ & $e_{\rm rms}$ &
 $\hspace{-0.3em}\langle{\rm SFR}\rangle$ & \hspace{1.5em}$A_{\rm IA}$
&  $10^2n_g$ & $e_{\rm rms}$ & \hspace{-0.3em}$\langle{\rm SFR}\rangle$ &\hspace{1.5em} $A_{\rm IA}$
&  $10^2n_g$ & $e_{\rm rms}$ & \hspace{-0.3em}$\langle{\rm SFR}\rangle$ &\hspace{1.5em} $A_{\rm IA}$ \\  \hline

0.3 &  $1.77$ & $0.23$ & $0.52$ & $0.08\pm 0.10$
	& $0.69$  & $0.28$ & $0.006$ & $2.29\pm 0.19$
	& $0.024$ & $0.33$ & $0.082$ & $19.23\pm 1.33$\\
	
0.5 &  $1.76$ & $0.24$ & $0.73$ & $0.11\pm 0.10$
	& $0.66$  & $0.29$ & $0.59$ & $2.49\pm 0.21$
	& $0.020$ & $0.33$ & $0.31$ & $16.82\pm 1.43$\\
	
0.7 &  $1.74$ & $0.25$ & $0.98$ & $0.19\pm 0.10$
	& $0.62$  & $0.30$ & $1.59$ & $2.48\pm 0.23$
	& $0.017$ & $0.33$ & $0.62$ & $17.52\pm 1.59$\\

1 &  $1.69$   & $0.26$ & $1.38$ & $0.04\pm 0.11$
	& $0.55$  & $0.29$ & $3.16$ & $2.52\pm 0.24$
	& $0.012$ & $0.33$ & $0.93$ & $19.22\pm 1.84$\\
	
1.5 & $1.50$  & $0.27$ & $2.43$ & $0.22\pm 0.13$
	& $0.41$  & $0.30$ & $6.31$ & $2.48\pm 0.28$
	& $0.007$ & $0.31$ & $0.89$ & $17.00\pm 2.32$\\
	
2   & $1.19$  & $0.29$ & $3.99$ & $0.28\pm 1.53$
	& $0.29$  & $0.30$ & $10.4$ & $1.53\pm 0.34$
	& $0.005$ & $0.32$ & $0.26$ & $7.63\pm 2.89$ \\ \hline \hline
\end{tabular}
\caption{Characteristics of the galaxy samples, selected from 
the stellar mass range in the TNG300 output at a given redshift. 
The stellar mass of galaxies is in units of $[\msun]$, and the number density of each sample is in units of $[10^{-2}(h^{-1}{\rm Mpc})^{-3}]$.
$\langle{\rm SFR}\rangle$ is the median star formation rate of galaxies in the sample, in units of $[M_\odot~ {\rm yr}^{-1}]$.
$e_{\rm rms}$ is the rms intrinsic ellipticities per component, for the definition of galaxy shape (Eq.~\ref{eq:ellipticity}). The IA coefficient, $A_{\rm IA}$, is estimated from the comparison of $P_{\delta E}$ and $P_{\delta\delta}$ in the three lowest $k$ bins in the range $k\simeq [0.1,0.3]~\hmpci$ (see text for details).
\label{tab:sample_mstar}}
\end{table}

\begin{table}
\centering
\scriptsize
\begin{tabular}{l|p{2.5em}p{1.5em}p{2.5em}cc|p{2.5em}p{1.5em}p{2.5em}cc} \hline\hline 
& \multicolumn{10}{c}{disks} \\
& \multicolumn{5}{c|}{$M_\star\in [10^9,10^{10}]\msun$} & \multicolumn{5}{c}{$M_\star\ge 10^{10}\msun$}
\\ 
$z$ & $10^2n_g$ & $e_{\rm rms}$ & \hspace{-0.3em}$\langle{\rm SFR}\rangle$ & $A_{\rm IA}$ & $A_{E_J}$ 
& $10^2n_g$ & $e_{\rm rms}$ & \hspace{-0.3em}$\langle{\rm SFR}\rangle$ & $A_{\rm IA}$ & $A_{E_J}$ 
\\ \hline
0.3 & 0.23 & 0.32 & 0.99 & $0.54 \pm 0.52$ & $-0.83 \pm 0.64$
	& 0.18 & 0.34 & 1.56 & $1.20 \pm 0.55$ & $0.33 \pm 0.72$\\
0.5 & 0.28 & 0.32 & 1.23 & $0.60 \pm 0.57$ & $-1.1 \pm 0.59$
	& 0.20 & 0.34 & 2.37 & $2.30 \pm 0.59$ & $0.78 \pm 0.71$\\
0.7 & 0.31 & 0.32 & 1.50 & $-0.50 \pm 0.56$ & $-1.38 \pm 0.54$
	& 0.20 & 0.34 & 3.27 & $2.35 \pm 0.64$ & $1.21 \pm 0.70$\\
1   & 0.33 & 0.32 & 1.90 & $-0.95\pm 0.60$ & $-2.58 \pm 0.54$
	& 0.20 & 0.34 & 4.89 & $2.93 \pm 0.69$ & $2.20 \pm 0.73$\\
1.5 & 0.31 & 0.32 & 2.93 & $-0.18 \pm 0.67$ & $-1.72 \pm 0.59$
	& 0.15 & 0.34 & 7.70 & $2.05 \pm 0.77$ & $3.08 \pm 0.80$\\
2   & 0.22 & 0.33 & 5.00 & $-0.71\pm 0.76$ & $-2.48 \pm 0.71$
 	& 0.12 & 0.34 & 10.8 & $0.95 \pm 0.88$ & $2.04 \pm 0.91$\\ \hline\hline
& \multicolumn{10}{c}{spheroids} \\
& \multicolumn{5}{c|}{$M_\star\in [10^9,10^{10}]\msun$} & \multicolumn{5}{c}{$M_\star\ge 10^{10}\msun$}
\\ 
$z$ & $10^2n_g$ & $e_{\rm rms}$ & \hspace{-0.3em}$\langle{\rm SFR}\rangle$ & $A_{\rm IA}$ & $A_{E_J}$
& $10^2n_g$ & $e_{\rm rms}$ & \hspace{-0.3em}$\langle{\rm SFR}\rangle$ & $A_{\rm IA}$ & $A_{E_J}$
\\ \hline
0.3 & 0.98 & 0.18 & 0.39 & $-0.35\pm 0.10$ & $-0.87\pm 0.30$
	& 0.34 & 0.25 & 0.00 & $4.02 \pm 0.25$ & $2.76 \pm 0.53$\\
0.5 & 0.92 & 0.20 & 0.58 & $-0.28\pm 0.12$ & $-0.70\pm 0.31$
	& 0.31 & 0.26 & 0.00 & $3.65 \pm 0.26$ & $1.73 \pm 0.56$ \\
0.7 & 0.89 & 0.21 & 0.80 & $-0.07\pm 0.13$ & $-0.02\pm 0.33$
	& 0.27 & 0.26 & 0.02 & $3.45 \pm 0.28$ & $2.34 \pm 0.58$\\
1   & 0.82 & 0.23 & 1.19 & $0.05 \pm 0.14$ & $0.16\pm 0.34$
	& 0.23 & 0.26 & 0.24 & $3.19 \pm 0.30$ & $1.75 \pm 0.66$\\
1.5 & 0.68 & 0.25 & 2.22 & $0.30 \pm 0.16$ & $-0.54\pm 0.39$
	& 0.16 & 0.26 & 1.97 & $3.28 \pm 0.36$ & $1.48 \pm 0.79$\\
2   & 0.60 & 0.27 & 3.58 & $0.34 \pm 0.19$ & $-0.18\pm 0.41$
 	& 0.09 & 0.26 & 8.55 & $1.48 \pm 0.44$ & $1.01 \pm 1.07$\\ \hline\hline
\end{tabular}
\caption{Similar to Table~\ref{tab:sample_mstar}, but for the sample of galaxies that are identified by either ``disks'' or ``spheroids'', respectively, selected from the parent sample with $M_\star\ge 10^9\msun$. To define the division, we defined galaxies with $\kappa_{\rm rot}>0.55$ as disk-like or rotation-supported galaxies, while we defined galaxies with $\kappa_{\rm rot}<0.45$ as spheroid-like or dispersion-dominated galaxies, where $\kappa_{\rm rot}$ is the ratio of the rotational kinetic energy to the total energy: $\kappa_{\rm rot}\equiv K_{\rm rot}/K$ (see the 
first paragraph in Section~\ref{subsec:gal_sample}).
\label{tab:sample_disk_spheroid}}
\end{table}

\begin{table}
\centering
\scriptsize
\begin{tabular}{l|p{2.em}p{1.em}p{2.3em}ccc|p{2.em}p{1.em}p{2.3em}ccc} \hline\hline 
& \multicolumn{12}{c}{centrals} \\
& \multicolumn{6}{c|}{$M_\star\in [10^9,10^{10}]\msun$} & \multicolumn{6}{c}{$M_\star\ge 10^{10}\msun$}
\\ 
$z$ & $10^2n_g$ & $e_{\rm rms}$ & ${\rm log}\langle M_{\rm h} \rangle$ & \hspace{-0.3em}$\langle{\rm SFR}\rangle$ & $A_{\rm IA}$ & $A_{E_J}$ 
& $10^2n_g$ & $e_{\rm rms}$ & ${\rm log}\langle M_{\rm h} \rangle$ & \hspace{-0.3em}$\langle{\rm SFR}\rangle$ & $A_{\rm IA}$ & $A_{E_J}$ 

\\ \hline
0.3 & 1.04 & 0.24 & 11.55 & 0.63 & $0.33 \pm 0.13$ & $0.22 \pm 0.30$
	& 0.46 & 0.28 & 12.56 & 0.085 & $3.91 \pm 0.25$ & $1.52 \pm 0.69$\\
0.5 & 1.06 & 0.25 & 11.56 & 0.85 & $0.41 \pm 0.14$ & $0.19 \pm 0.30$
	& 0.45 & 0.29 & 12.53 & 0.893 & $3.71 \pm 0.27$ & $2.94 \pm 0.56$\\
0.7 & 1.07 & 0.26 & 11.57 & 1.07 & $0.50 \pm 0.14$ & $0.34 \pm 0.29$
	& 0.43 & 0.29 & 12.50 & 1.868 & $3.70 \pm 0.28$ & $3.23 \pm 0.49$\\
1   & 1.09 & 0.26 & 11.57 & 1.47 & $0.32 \pm 0.15$ & $0.14 \pm 0.30$
	& 0.39 & 0.30 & 12.45 & 3.439 & $4.00 \pm 0.30$ & $2.78 \pm 0.47$\\
1.5 & 1.03 & 0.28 & 11.57 & 2.51 & $0.39 \pm 0.16$ & $-0.33 \pm 0.32$
	& 0.31 & 0.30 & 12.37 & 6.435 & $3.40 \pm 0.32$ & $3.00 \pm 0.46$\\
2   & 0.87 & 0.29 & 11.57 & 4.08 & $0.59 \pm 0.19$ & $-0.42 \pm 0.35$
 	& 0.23 & 0.30 & 12.30 & 10.424 & $2.42 \pm 0.38$ & $3.60 \pm 0.45$\\ \hline\hline
& \multicolumn{12}{c}{satellites} \\
& \multicolumn{6}{c|}{$M_\star\in [10^9,10^{10}]\msun$} & \multicolumn{6}{c}{$M_\star\ge 10^{10}\msun$}
\\ 
$z$ & $10^2n_g$ & $e_{\rm rms}$ & --- & $\langle{\rm SFR}\rangle$ & $A_{\rm IA}$ & $A_{E_J}$ 
& $10^2n_g$ & $e_{\rm rms}$ & --- & $\langle{\rm SFR}\rangle$ & $A_{\rm IA}$ & $A_{E_J}$ 
\\ \hline
0.3 & 0.74 & 0.21 & --- & 0.31 & $-0.25 \pm 0.13$ & $-1.85 \pm 0.35$
	& 0.25 & 0.27 & --- & 0.00 & $0.95 \pm 0.31$ & $-1.07 \pm 1.27$\\
0.5 & 0.70 & 0.22 & --- & 0.52 & $-0.33 \pm 0.15$ & $-2.01 \pm 0.37$
	& 0.23 & 0.28 & --- & 0.065 & $1.31 \pm 0.33$ & $0.19 \pm 0.94$\\
0.7 & 0.67 & 0.24 & --- & 0.77 & $-0.31 \pm 0.16$ & $-1.16 \pm 0.38$
	& 0.20 & 0.29 & --- & 0.76 & $1.11 \pm 0.36$ & $-1.28 \pm 0.77$\\
1   & 0.61 & 0.25 & --- & 1.17 & $-0.46 \pm 0.18$ & $-1.55 \pm 0.41$
	& 0.17 & 0.30 & --- & 2.24 & $0.24 \pm 0.42$ & $0.96 \pm 0.67$\\
1.5 & 0.47 & 0.27 & --- & 2.21 & $-0.19 \pm 0.23$ & $-0.93 \pm 0.49$
	& 0.11 & 0.30 & --- & 5.38 & $0.89 \pm 0.54$ & $-0.13 \pm 0.63$\\
2   & 0.33 & 0.28 & --- & 3.72 & $-0.55 \pm 0.30$ & $-1.29 \pm 0.58$
 	& 0.063 & 0.31 & --- & 9.73 & $-1.35 \pm 0.75$ & $-0.27 \pm 0.62$\\
\hline\hline
\end{tabular}
\caption{Similar to the previous table, but for the galaxy samples that are either centrals or satellites in their host halos, selected from the parent sample with $M_\star\ge 10^9\msun$. $\log\langle M_{\rm h}\rangle$ is the log of the average mass of host halos for each sample, and the halo mass $M_{\rm h}$ is in unit of $\msun$.
\label{tab:sample_central}}
\end{table}

\subsection{Ellipticity measurements}
\label{subsec:e_measurement}

In this section, we describe how we estimate the ellipticity of individual galaxies in the simulation. In an actual observation, we can only observe the projected shape of each galaxy on the sky. Based on this consideration, we characterize the galaxy shape by the ellipticity parameter, $\epsilon_i$:
\begin{equation}
    \binom{\epsilon_+}{\epsilon_\times}=\frac{a^2-b^2}{a^2+b^2}\binom{\cos 2\theta_{\rm P}}{\sin 2\theta_{\rm P}},
\label{eq:ellipticity}
\end{equation}
where $a$ and $b$ are the semi-major and semi-minor axes when we approximate the projected galaxy shape by an ellipse, and $\theta_{\rm P}$ is the position angle of the major axis with respect to the 1st coordinate axis on the sky. 
Thus the galaxy ellipticity is a coordinate-dependent quantity. The ellipticity field forms a spin-2 field, since the ellipse stays unchanged under a $180^{\circ}$ rotation of the coordinate axes.

In simulations, there exist two ways to characterize the galaxy ellipticities. The first one is to project the 3D galaxy spheroid, described by the inertia tensor of stellar mass, along the line of sight \cite{2015MNRAS.448.3522T}.
The other one is to use the spin of galaxy, which is characterized by the stellar angular momentum \cite{2015MNRAS.452.3369C}.
These two methods are complementary to each other. The implementation of these two methods in IA studies is mainly motivated by the different theoretical mechanisms that are thought to be responsible for regulating the observed ellipticities of different galaxy populations.

For disk galaxies, they are rotation dominated; the angular momentum vector direction with respect to the line of sight mainly determines their observed ellipticities.
For spheroidal galaxies, they are mainly random motion dominated; the matter distribution within the galaxy projected on the sky gives the observed ellipticity within a halo. 
The angular momentum of disk galaxies is thought to originate from the tidal torque, while the mass distribution of spheroids is more dictated by the tidal stretching. Both of them will also be modulated by a complex combination of galaxy formation physics such as mergers and baryonic physics. We shall discuss more on the theoretical formalisms in Section~\ref{sec:formalism}.

\subsubsection{Ellipticity measured with inertia tensor}
\label{subsubsec:e_inertia}
To obtain the ellipticity of galaxies, one way is to model the 3D stellar matter distribution as an ellipsoid and then project it onto the $x_{1}x_{2}$ plane of simulation coordinates (assuming the line-of-sight direction to be along the $x_3$ axis).
This results in an ellipse that characterizes the galaxy shape similar to what we observe. In this work, we adopt the reduced mass inertia tensor to describe the 3D spheroid \cite{2012MNRAS.420.3303B,2015MNRAS.448.3522T,Osatoetal2018,2020arXiv200412579K}, which gives more weight to the stellar particles near the center of galaxy. The reduced inertia tensor is defined as
\begin{equation}
    I_{ij}=\frac{\sum_{n} m_n \frac{x_{ni} x_{nj}}{r_n^2}}{\sum_{n} m_n}  
\end{equation}
where $m_n$ is the mass of the $n$th stellar particle within the subhalo, $x_{ni}$, $x_{nj} (i,j=1,2,3)$ are the position coordinates of this particle with respect to the centre of galaxy. The inertia tensor is calculated iteratively, which, according to Ref.~\cite{2015MNRAS.448.3522T}, must be adopted to have a reliable measurement of the reduced inertia tensor. {In the iterative scheme, we first obtain the axis ratio by the using the spherical weighting, i.e. $r^2_{n}=\sum_i x^2_{ni}$. Then we iteratively weight each particle by using the triaxial weighting, i.e. $r^{2}_n=\left(\bx_n \cdot \bm{e}_a \right)^2 + \left( \frac{\bx_n \cdot \bm{e}_b}{s}\right)^2 + \left(\frac{\bx_n \cdot \bm{e}_c}{q}\right)^2$, where $\bm{e}_a$, $\bm{e}_b$, and $\bm{e}_c$ are the eigen vector of the inertia tensor spheroid, $a>b>c$ are the corresponding eigen axis, $s\equiv c/a$ and $q\equiv b/a$. By keeping the major axis $a$ constant, we repeat the triaxial weighting until $s$ and $q$ converge to 1\%.} Please refer to Refs.~\cite{2012MNRAS.420.3303B} and \cite{2015MNRAS.448.3522T} for more details on the comparison of different inertia tensor definitions. 

As the $x_3$-axis is taken as the light-of-sight direction, the ellipticities of the projected ellipse are given as
\begin{equation}
\label{eq:e_inertia}
    \epsilon_+\equiv\frac{I_{11}-I_{22}}{I_{11}+I_{22}}, \epsilon_\times \equiv\frac{2I_{12}}{I_{11}+I_{22}}.
\end{equation}
This ellipticity measurement in simulation has the advantage that it resembles the galaxy ellipticities measured from the quadrupole moments of the surface brightness of the galaxy images \citep{1995ApJ...449..460K,2005MNRAS.361.1287M}.
The IA correlation or power spectrum is measurable only in a statistical sense, by correlating galaxy ellipticities themselves or cross-correlating galaxy ellipticities with positions of galaxies or matter. The IA correlation amplitudes vary with 
the estimator of individual galaxy shapes. 
For example, if we employ the unreduced inertia tensor, i.e., do not use the radial weight $1/r^2$ in the inertia tensor definition, it alters the overall amplitude of the IA correlations, and causes a slight modification in the scale dependence at nonlinear scales 
\cite{2016MNRAS.457.2301S,2020arXiv200412579K}. However, as studied in Ref.~\cite{2020arXiv200412579K}, 
the signal-to-noise ratio for a measurement of the IA power spectrum is almost unchanged. This means that, for the IA measurements, the accuracy of individual galaxy shapes is not critical, unlike weak lensing. Instead one has to treat the overall coefficient in the IA power spectrum ($A_{\rm IA}$ as defined below) as a free parameter, very similarly to what is done for the linear bias parameter ($b_g$) in the analysis of the galaxy density power spectrum. In Appendix~\ref{sec:varying_e_def} we explicitly study how the IA power spectrum varies with the different definitions of galaxy ellipticities for the same sample of galaxies.

\subsubsection{Ellipticity measured with angular momentum}
\label{subsubsec:e_J}

An alternative way to characterize the galaxy shape is
to use the galaxy angular momentum as a proxy of the galaxy
ellipticity. This provides us a measurement of the spin alignment of galaxies that is in a similar form to that of the shape alignment. If the observed ellipticity of galaxy is indeed determined by the angular momentum, we expect to see a very similar IA signal of this ellipticity measured with spin to the ellipticity measured with inertia tensor.
Assuming that rotation-supported galaxies form a thin, circular-shaped disk, their observed ellipticity can be estimated from the inclination of the disk with respect to the line-of-sight direction. 
We use the method in Ref.~\cite{2001ApJ...559..552C} to define the observed ellipticity from the orientation of the angular momentum vector with respect to the line-of-sight direction (the $x_3$-axis direction in this work):
\begin{equation}
\label{eq:e_J}
    \epsilon_+\equiv-\frac{{\hat L_1^2}-{\hat L_2^2}}{1+{\hat L_3^2}}, 
    \epsilon_\times\equiv-\frac{2{\hat L_1^2}{\hat L_2^2}}{1+{\hat L_3^2}},
\end{equation}
where ${\hat L_i}$ ($i=1$, $2$, $3$) are the components of the unit angular momentum vector of the galaxy with respect to the center, which is calculated by summing up the angular momentum of all stellar particles within the galaxy, i.e.,
\begin{equation}
 {\hat L_i^2}\equiv \frac{1}{|\bm{L}|^2}\sum_n L^2_{i,n}.
\end{equation}
Here $L_{i,n}$ is the angular momentum component of the $n$-th particle along the $x_i$ direction, and $\bm{L}=\sum_n m_n \bm{x}_n\times\bm{v}_n$ is the galaxy angular momentum vector, where $\bm{x}_n$ and $\bm{v}_n$ are the position vector and velocity vector of the $n$-th stellar particle relative to the centre of mass of the galaxy. $\epsilon_i=0$ corresponds to a round ellipse, when $\bm{L}=L\hat{\bm{x}}_3$, or equivalently
when the disk is seen from the face-on view. $|\epsilon_i|=1$ corresponds to an elongated stick, when the disk is seen from the edge-on view. However, in practice, the projected ellipticity of an axis-symmetric disk galaxy depends on the thickness of the disk 
\cite{2013MNRAS.431..477J,2013MNRAS.436..819J}, 
which will decrease the ellipticities measured above. Unfortunately, the disk thickness in TNG300 is overestimated due to the limited scale resolution. 
Thus we ignore the disk thickness in this work. In addition, an actual galaxy would have complex structures in its disk such as star-forming knots, HII regions and dust layers, and these would give a non-trivial definition of the ``observed'' ellipticity on individual galaxy basis. However, as we discussed above, as long as these fine structures appear randomly between different galaxies, they can be regarded as a higher-order effect altering only the linear IA coefficient and just add statistical noise to the IA correlation measurements. We believe that a simple characterization of the galaxy shape by the angular momentum vector can capture the main effect of IA correlations for disk galaxies. 

Our definition of galaxy ellipticities, following other works, is in units of $(a^2-b^2)/(a^2+b^2)$ in terms of the major and minor axes lengths, $a,b$, of an ellipse. In order to make the ellipticities consistent with the weak lensing shear, which is given as $(a-b)/(a+b)$, we define the IA shear, $\gamma_{+,\times}=\epsilon_{+,\times}/(2{\cal R})$,
where $\mathcal{R} \equiv 1-\langle \epsilon_i^2\rangle$ is the responsivity \cite{2002AJ....123..583B}
and $\langle \epsilon_i^2\rangle\equiv \frac{1}{N_{g}}\sum_g \epsilon_{i,g}^2$ is the ellipticity variance per component for a given galaxy sample. 
The conversion from ellipticity to shear enables us to make a direct comparison with the IA theory predictions and makes it easier to compare with the previous works.
However, exactly speaking, this conversion is not necessary for the IA study, as long as the linear IA coefficient is treated as a free parameter. This is somewhat a convention used in the literature, mainly because the IA effect has often been discussed as a contaminating effect to 
weak lensing.

\subsection{Measurements of the three-dimensional IA power spectrum}
\label{subsec:measurement_method}

Here we briefly review the method to measure the three-dimensional IA power spectrum based on the $E/B$ decomposition introduced in Ref.~\cite{2020arXiv200412579K}.
Throughout this paper, we adopt the plane-parallel approximation and ignore the redshift-space distortion for simplicity.
As shown in Ref.~\cite{2020arXiv200412579K},
the spin-2 nature of the shear field (Eq.~\ref{eq:ellipticity})
indicates that we can perform the $E/B$-mode decomposition, as had been widely done in CMB polarization and weak lensing analyses. The scalar gravitational potential induces only the $E$-mode (curl-free) shear, while the $B$-mode (divergence-free) shear in the linear regime can only be generated by the non-vanishing systematics. 
The $E/B$-mode decomposition of galaxy shear in Fourier space are defined as
\begin{align}
	\gamma_E(\bm{k}) &=  \gamma_{+}(\bm{k}) \cos{2\phi_{\bm{k}}}	+  \gamma_{\times}(\bm{k}) \sin{2\phi_{\bm{k}}},
	\label{eq:gammaE}\\
	\gamma_B(\bm{k}) &= - \gamma_{+}(\bm{k}) \sin{2\phi_{\bm{k}}} +  \gamma_{\times}(\bm{k}) \cos{2\phi_{\bm{k}}},
	\label{eq:gammaB}
\end{align}
where {the three-dimensional wave vector is}
$\bm{k}=k(\sqrt{1-\mu^2}\cos\phi_{\bm{k}},\sqrt{1-\mu^2}\sin\phi_{\bm{k}},\mu)$; $\mu$ is the cosine angle between 
$\bm{k}$ and the {$x_3$}-axis direction, and $\phi_{\bm{k}}$ is the azimuthal angle between $\bm{k}$ and the {$x_1$} axis in Fourier space.
Following Ref.~\cite{2020arXiv200412579K}, we can define the {\it three-dimensional} IA power spectra of galaxies as
\begin{align}
    \langle\gamma_E(\bm{k})\gamma_E(\bm{k'})\rangle &\equiv (2\pi)^3\delta_D(\bm{k}+\bm{k'})P_{EE}(\bm{k}),
    \label{eq:ps_ee}\\
    \langle\gamma_E(\bm{k})\delta_m(\bm{k'})\rangle &\equiv 
    (2\pi)^3\delta_D(\bm{k}+\bm{k'})P_{\delta E}(\bm{k}),
    \label{eq:ps_me}\\
    \langle\gamma_E(\bm{k})\delta_g(\bm{k'})\rangle &\equiv 
    (2\pi)^3\delta_D(\bm{k}+\bm{k'})P_{g E}(\bm{k}),
    \label{eq:ps_ge}
\end{align}
where $\delta_D(\bm{k}+\bm{k'})$ is the 3D Dirac delta function.  
$P_{E E}$ is the auto power spectrum of the $E$-mode shear field, i.e., the II signal, which comes from the intrinsic alignment between galaxies lying in the same large scale structure. $P_{\delta E}$ is the cross power spectrum between mass density field and $E$-mode shear, and $P_{g E}$ is the cross power spectrum between galaxy overdensity field $\delta_g$ and $E$-mode shear field. 
$P_{\delta E}$ and $P_{g E}$ are also used as an indicator of the strength of the GI signal, which is one of the physical contaminating effect to cosmic shear measurements. The foreground large-scale structure causes aligned shapes of galaxies at the same redshift, and it also gives rise to a weak lensing shearing of background galaxy shapes, causing correlations between foreground galaxy shapes and background galaxy shapes \citep{HirataSeljak2004}, which is known as the GI signal.
In this paper, however, we study the IA effect as a ``signal'',  assuming that both spectroscopic and imaging surveys observing the same region of sky are available, which enables one to measure the auto and cross correlations between the positions and shapes in the {\it same} large-scale structure (in the very similar redshift range). 
We do {\it not} consider correlations between foreground and background galaxies (for their shapes and positions) which are 
a contamination to weak lensing measurements. We should note that the 3D IA power spectrum carries all the information of IA effect at the level of two-point statistics.

The spectra $\langle\gamma_B\delta_{m}\rangle$, $\langle\gamma_B\delta_{g}\rangle$, and $\langle\gamma_B\gamma_E\rangle$ should all vanish
due to the statistical parity invariance. The auto spectrum $\langle\gamma_B\gamma_B\rangle$ has non-zero values, arising from the intrinsic shape noise and the subtle contribution of nonlinear IA effects \citep[see][for details]{2020arXiv200412579K}. Thus we focus on the $E$-mode IA power spectrum in this work.

Below we briefly describe how we measure the IA power spectrum in the simulation. {The density and shear fields are obtained by assigning the mass elements/shear of galaxies to a $1024^3$ uniform Cartesian mesh using the nearest grid point (NGP) interpolation} scheme. For the details of the calculation procedure, we refer the readers to the Appendix~A in Ref.~\cite{2020arXiv200412579K}. 
Then we apply the fast Fourier transform (FFT) to calculate the power per mode as defined by Eqs.~(\ref{eq:ps_ee})--(\ref{eq:ps_ge}).
To extend the reliable dynamical range of the measured power spectrum in the simulation, we use the `self-folding' trick \citep{1998ApJ...499...20J,Valageas11a,2018MNRAS.475..676S} to have measurements of the power spectrum at $k\ge 1.96~\hmpci$ up to a folding factor of $2^5(=32)$. This enables us to probe the power spectrum over an effective range of $0.1<k<60\mpc$ for TNG300.

We further define the multipole moments of the IA power spectrum as
\begin{align}
    P^{(\ell)}(k)=\frac{2\ell+1}{2}\int^1_{-1}d\mu \mathcal{L}_\ell(\mu)P(k,\mu), 
\end{align}
where $P(k,\mu)$ is one of the power spectra defined in Eqs. (\ref{eq:ps_ee})--(\ref{eq:ps_ge}), {\bm{$\mu$} is $cos$ of the angle between $\bm{k}$ and the line of sight direction,} $\mathcal{L}_\ell$ is the Legendre polynomial. $P^{(0)}$ is the monopole component, $P^{(2)}$ is the quadrupole component, and $P^{(4)}$ is the hexadecapole component.

\section{Theoretical Models}
\label{sec:formalism}
In this section, we briefly review two analytical models for IA:
linear/nonlinear alignment model and quadratic alignment model. The following formalism has been presented by previous works (see Ref.~\cite{HirataSeljak2004, Blazeketal2015} for more details).
\subsection{Linear/Nonlinear alignment model}
\label{subsec:NLA}

The linear alignment (hereafter LA) model \cite{Catelanetal2001,HirataSeljak2004} predicts that shapes of galaxies originate from the tidal field of gravitational potential. It is usually thought that the LA model can well describe correlations between shapes of elliptical or early-type galaxies. In this model, the shape of each galaxy is assumed to be proportional to the surrounding tidal field as
\begin{align}
\gamma^{I}_{(+,\times)}=-\frac{C_1}{4\pi G} (\nabla_1^2-\nabla_2^2, 2\nabla_1\nabla_2)S[\Psi_{\rm p}],  
\end{align}
where $\nabla_1\equiv \partial/\partial x_1$ and so on, 
$\Psi_{\rm p}$ is the gravitational potential field at the time of the galaxy formation, $S$ is the smoothing filter of the matter field which is needed to filter out the nonlinear scale relevant for galaxy formation/physics. 
If a galaxy is formed at a high redshift in the matter dominated era, the potential field is constant in time and is equivalent to the primordial potential (therefore the subscript ``p'' stands for ``primordial'').
$C_1$ is the normalization parameter, and $C_1>0$ means that the galaxy is aligned with the stretching direction of the tidal field. We would like to point out that, the time evolution of IA caused by the galaxy advection, i.e. the peculiar motion of galaxies, is not included in the above formalism \citep[see][]{2018JCAP...07..030S}. The galaxy position that we observe at a certain time differs from the position at the time of formation due to its peculiar motion. As a result, the IA that we observe varies in the nonlinear regime.
The above equation holds only on large scales, much greater than scales of galaxy physics, and therefore is valid only in the statistical sense. In other words, on individual galaxy basis, the intrinsic shapes dominate over the cosmological signal or we cannot distinguish the cosmological LA contribution from the observed shape \citep[see][]{2020arXiv200412579K}.
The coefficient $C_1$ varies with types of galaxies, and the dependences of the IA signal on galaxy properties, such as luminosity, color, morphology, redshift, etc. are absorbed into the $C_1$ parameter. In the following we assume that the IA effect is imprinted in the matter dominated era. 

The potential field is related to the matter density fluctuation field via the Poisson equation as
\begin{align}
    \Psi_{\rm p}(\bm k, z_{\rm IA})=-4\pi G\bar{\rho}_{\rm m0}(1+z_{\rm IA})k^{-2}\delta({\bm k, z_{\rm IA}}),
\end{align}
where $\bar{\rho}_{\rm m0}$ is the mean matter density today, and $z_{\rm IA}$ is the time when the galaxies are formed or equivalently when the IA alignment is imprinted. Thus the IA shear can be expressed as 
\begin{align}
    \gamma^{I}_{(+,\times)}(\bm{k},z)=-A_{\rm IA}C_1\frac{\bar{\rho}_{\rm m0}}{D(z)}f_{(+,\times)}\delta({\bm k, z}),
\end{align}
where $f_+\equiv (1-\mu^2)\cos2\phi_{\bm{k}}$, $f_\times \equiv (1-\mu^2)\sin2\phi_{\bm{k}}$, and $D(z)$ is the linear growth factor normalized as $D(z=0) = 1$. 
Note that the field $\delta(\bm{k},z)/D(z)$, with the factor of $1/D(z)$, converts it to the density fluctuation field in the matter dominated era because $\delta(\bm{k},z)\propto D(z)$ in the linear regime.

Hence the LA model predicts that the IA power spectra are given in terms of the underlying linear matter power spectrum as
\begin{align}
    &P_{\delta E}(k,\mu)=-A_{\rm IA}C_1\rho_{\rm cr0}\frac{\Omega_{\rm m}}{D(z)}(1-\mu^2)P_{\delta\delta}^{\rm lin}(k,z),
    \label{eq:A_IA_dE}\\
    &P_{EE}(k,\mu)=\left[A_{\rm IA}C_1\rho_{\rm cr0}\frac{\Omega_{\rm m}}{D(z)}\right]^2(1-\mu^2)^2P_{\delta\delta}^{\rm lin}(k,z),
    \label{eq:A_IA_EE}\\
    &P_{g E}(k,\mu)=-A_{\rm IA}C_1\rho_{\rm cr0}\frac{\Omega_{\rm m}}{D(z)} b_g(1-\mu^2) P_{\delta \delta}^{\rm lin}(k,z),
    \label{eq:A_IA_gE}
\end{align}
where $\rho_{\rm cr0}$ is the critical density today, and $P^{\rm lin}_{\delta\delta}(k,z)$ is the linear matter power spectrum at redshift $z$. Here we assume that the number density fluctuation field of galaxies is related to the matter density field, via a constant linear bias parameter $b_g$, as $\delta_g=b_g\delta$, which is a good approximation on large scales in the linear regime.

Following the convention in Ref.~\cite{Joachimi2011}, throughout this paper we employ $C_1\rho_{\rm cr0}=0.0134$, and 
introduce $A_{\rm IA}$, instead of $C_1$, to denote the linear IA coefficient for different types of galaxies. Note that $A_{\rm IA}$ is dimension-less.
In this paper, we ignore the redshift-space distortion due to peculiar velocities of matter or galaxies. The above equations clearly show that the IA power spectra are a two-dimensional function, given as a function of $(k,\mu)$, and the amplitudes depend on $\mu$, in addition to the length of $\bm{k}$, i.e., $k$.
Thus the cross power spectra, $P_{\delta E}$ and $P_{gE}$, have angular modulations up to $\mu^2$, or equivalently have non-zero monopole and quadrupole moments. The auto spectrum $P_{EE}$ has angular modulations up to $\mu^4$, i.e., non-zero monopole, quadrupole and hexadecapole moments.

We expect that the LA model gives an accurate description of the IA effect on large scales in the linear regime, for the $\Lambda$CDM model with the adiabatic initial conditions. Ref.~\cite{2007NJPh....9..444B} proposed an empirical model to apply the IA model to nonlinear scales by 
replacing the linear matter power spectrum with the nonlinear one. 
This model is known as the non-linear alignment (NLA) model. However, this model suffers from the inconsistencies in the physical origin and fails when it goes to smaller scales \citep{2011JCAP...05..010B,Blazeketal2017}.

\subsection{Quadratic alignment model}
\label{subsec:QA}
As we described above, we deduce that shapes of star-forming or spiral galaxies are mainly determined by the angular momentum, rather than the tidal field. The so-called ``quadratic alignment model'', which is developed based on the tidal torque theory for the origin of angular momentum \cite{1969ApJ...155..393P,1970Ap......6..320D,1984ApJ...286...38W,1996MNRAS.282..436C,Catelanetal2001,HirataSeljak2004},
predicts that the shape of such galaxies arises from the torque of the tidal field
\begin{align}
    \gamma^{I}_{(+,\times)}=C_2(T^2_{1 i}-T^2_{2i}, 2T_{1 i }T_{2 i}),
\end{align}
where the tidal tensor is 
\begin{align}
    T_{ij}=\frac{1}{4\pi G}(\nabla_i\nabla_j-\frac{1}{3}\delta^K_{ij}\nabla^2)S[\Psi_{\rm p}].
\end{align}
Comparing with the LA/NLA, where the tidal field contributes to the galaxy shear in first order, the quadratic alignment arises from the second-order contributions of the tidal field. Hence this model predicts that the IA power spectrum arising from the quadratic alignment does not have the linear limit by definition. More explicitly, if this quadratic alignment model is correct, the ratio of the IA cross power spectrum to the matter power spectrum, $P_{\delta E_J}/P_{\delta\delta}$, should vanish at the limit of $k\rightarrow 0$: $\lim_{k\rightarrow 0}P_{\delta E_J}/P_{\delta\delta} \rightarrow 0$. However, note that non-linear evolution can introduce a correlation even at finite $k$ scales in the  linear scales as discussed in Refs.~\cite{2002astro.ph..5512H} and \cite{2016PhRvD..94b2002B}.
In the following, we denote the $E$ mode of the angular momentum shear as $E_J$.
We will use the simulation to check whether this asymptotic behavior holds for galaxy shapes seen in the TNG300 hydro simulations.

\section{Results}
\label{sec:result}

In this section, we first show the IA power spectrum measured in TNG300, discussing its stellar mass dependence and redshift evolution. In Section~\ref{subsec:IA_fixng}, we show the IA for galaxy samples that are defined by selecting galaxies from the ranked list of stellar mass ($M_\star$) or SFR from its largest one until the number density matches a target density of $n_g=10^{-4}~(h^{-1}{\rm Mpc})^{-3}$, which is a typical number density for the existing and upcoming spectroscopic galaxy surveys \citep{2013AJ....145...10D,2016AJ....151...44D,2014PASJ...66R...1T,2016arXiv161100036D}.
The morphology (disk versus spheroid) and environmental (central versus satellite) dependences of IA are shown in Sections~\ref{subsec:IA_morph} and \ref{subsec:IA_censat}. Since we only study the monopole component of the power spectrum, such as $P^{(0)}_{\delta E}$ and $P^{(0)}_{\delta E_J}$, for simplicity, we are going to omit the superscript of $(0)$ for notational simplicity in the following.

Tables~\ref{tab:sample_mstar}--\ref{tab:sample_central} summarizes 
the sample properties, including the number density, rms ellipticity ($e_{\rm rms}$), median SFR, and the linear IA coefficient $A_{\rm IA}$. Here we estimate $A_{\rm IA}$ for each sample by performing a fitting between the IA cross power spectrum, $P_{\delta E}$, and $P_{\delta\delta}$ in the three lowest $k$ bins, $k\simeq [0.1,0.3]~\hmpci$, assuming the relation of $P_{\delta E}\propto A_{\rm IA}P_{\delta\delta}$ in Eq.~(\ref{eq:A_IA_dE}). The galaxy sample having a larger $A_{\rm IA}$ means that the galaxy shapes have a stronger alignment with the tidal field. For some stellar-mass samples, 
the significance of $A_{\rm IA}$ reaches about $A_{\rm IA}/\sigma_{A_{\rm IA}}\sim 20$, i.e., $20\sigma$ detection. These results can be compared with the actual measurements for the BOSS LOWZ galaxies shown in Table~2 of Ref.~\cite{2015MNRAS.450.2195S}. The LOWZ galaxies are approximately close to a massive, stellar mass-limited sample. The significance of $A_{\rm IA}$ for TNG300 is comparable or even better than the LOWZ measurements, even if the TNG300 volume is much smaller than that of the BOSS; roughly $(205~h^{-1}{\rm Mpc})^3$ vs. $(1~h^{-1}{\rm Gpc})^3$.  This reflects the power of the 3D IA power spectrum method, while Ref.~\cite{2015MNRAS.450.2195S} uses the projected correlation function (see Ref.~\cite{2020arXiv200412579K} for the similar discussion). However, we want to emphasize that a quantitative comparison between the simulation and observation is hampered by the different scale ranges studied, not exactly the same samples, and the neglect of the RSD effect in our work, and thus a more proper comparison requires a careful work taking into account these effects, which is
beyond the scope of this work.

The table also gives the number density and the rms intrinsic ellipticities $e_{\rm rms}$, because a measurement of the IA power spectrum is affected by the shape noise scatter that is proportional to $e_{\rm rms}^2/n_g$. For disk-like galaxies, which are rotation-supported galaxies for their kinematics, we also give the fitting results for the angular-momentum {induced} IA effect, denoted as $A_{E_J}$. Although the angular momentum IA has a different mechanism from the tidal alignment model, we simply use the same relation, Eq.~(\ref{eq:A_IA_dE}), to estimate the $A_{E_J}$ coefficient in order to make it easier to compare the results. In the following we give more detailed discussion for each galaxy sample. 

\subsection{IA power spectrum of galaxies in TNG300: {stellar mass dependence and redshift evolution}}
\label{subsec:mstar_z}

\begin{figure}[!htb]
\centering 
\includegraphics[width=.99\textwidth]{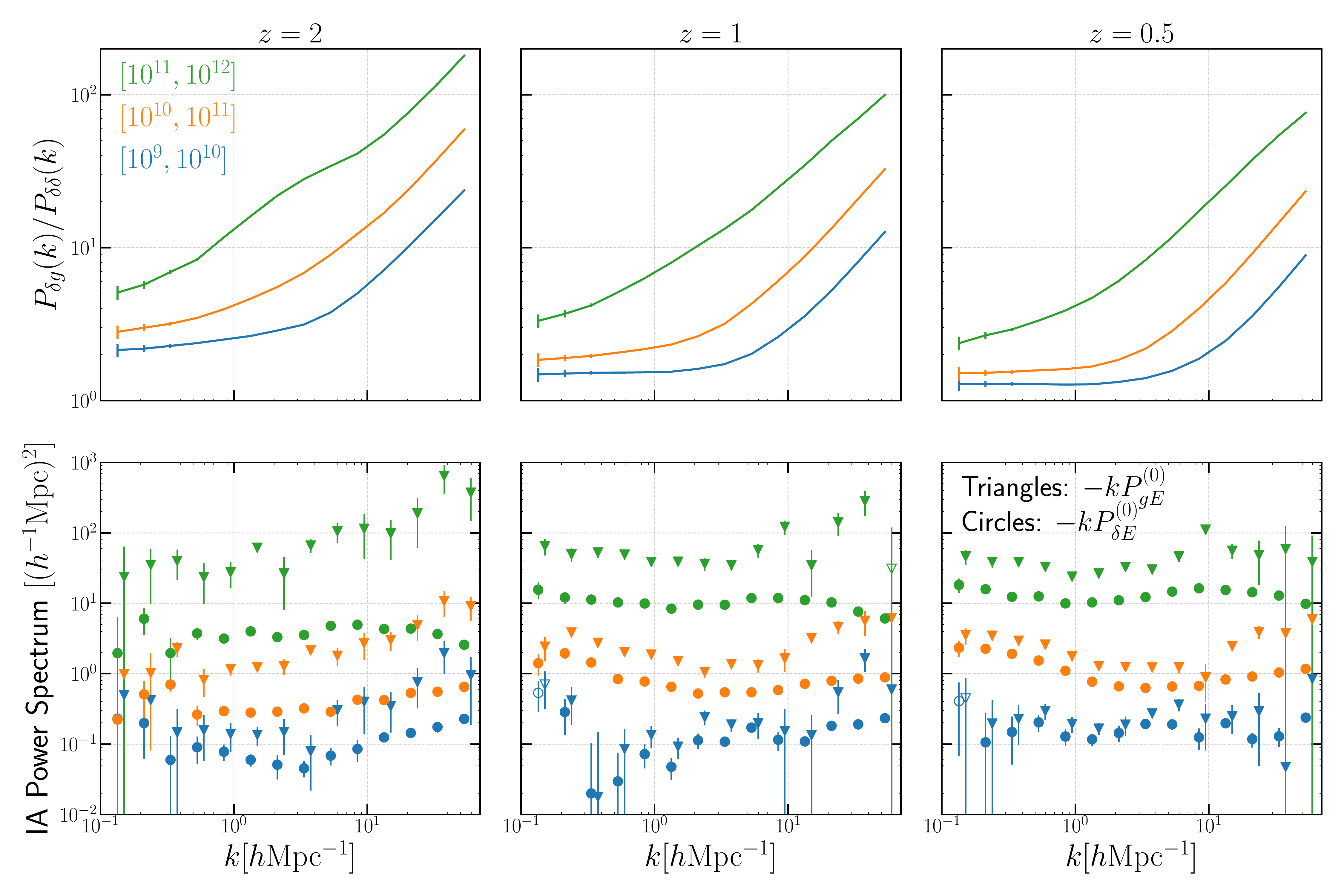}
\caption{{\it Upper panels}: the ratio of the galaxy-matter cross power spectrum and the matter power spectrum, $P_{\delta g}/P_{\delta\delta}$. The small-$k$ limit of this ratio corresponds to the linear bias parameter $b_1$ for each sample. The green, orange, and blue lines are galaxies in the stellar mass ranges of $10^{11}<M_\star/\msun<10^{12}$, $10^{10}<M_\star/\msun<10^{11}$, and $10^{9}<M_\star/\msun<10^{10}$, respectively. {$P_{\delta g}/P_{\delta\delta}$ ratio increases with $M_\star$, and deviates from a constant and rises up towards large $k$.} {\it Lower panels}: the cross power spectrum of IA shear with the mass density fluctuation field, $P_{\delta E}$ (circles),
or the galaxy overdensity field, $P_{gE}$ (triangles), respectively,for galaxies in three stellar mass bins at different redshifts. For illustrative purpose, we multiply the power spectra by $-k$ so that the range of the power spectrum is in a narrower range of $y$-axis. The negative values are shown using open symbols. The strength of the IA power spectrum of galaxies increases with the stellar mass and stays roughly unchanged with redshift. Due to the galaxy bias, the amplitude of $P_{gE}$ is higher than $P_{\delta E}$.}
\label{fig:Pk_E0_mstarz} 
\end{figure}
\begin{figure}[!htb]
\centering 
\includegraphics[width=.9\textwidth]{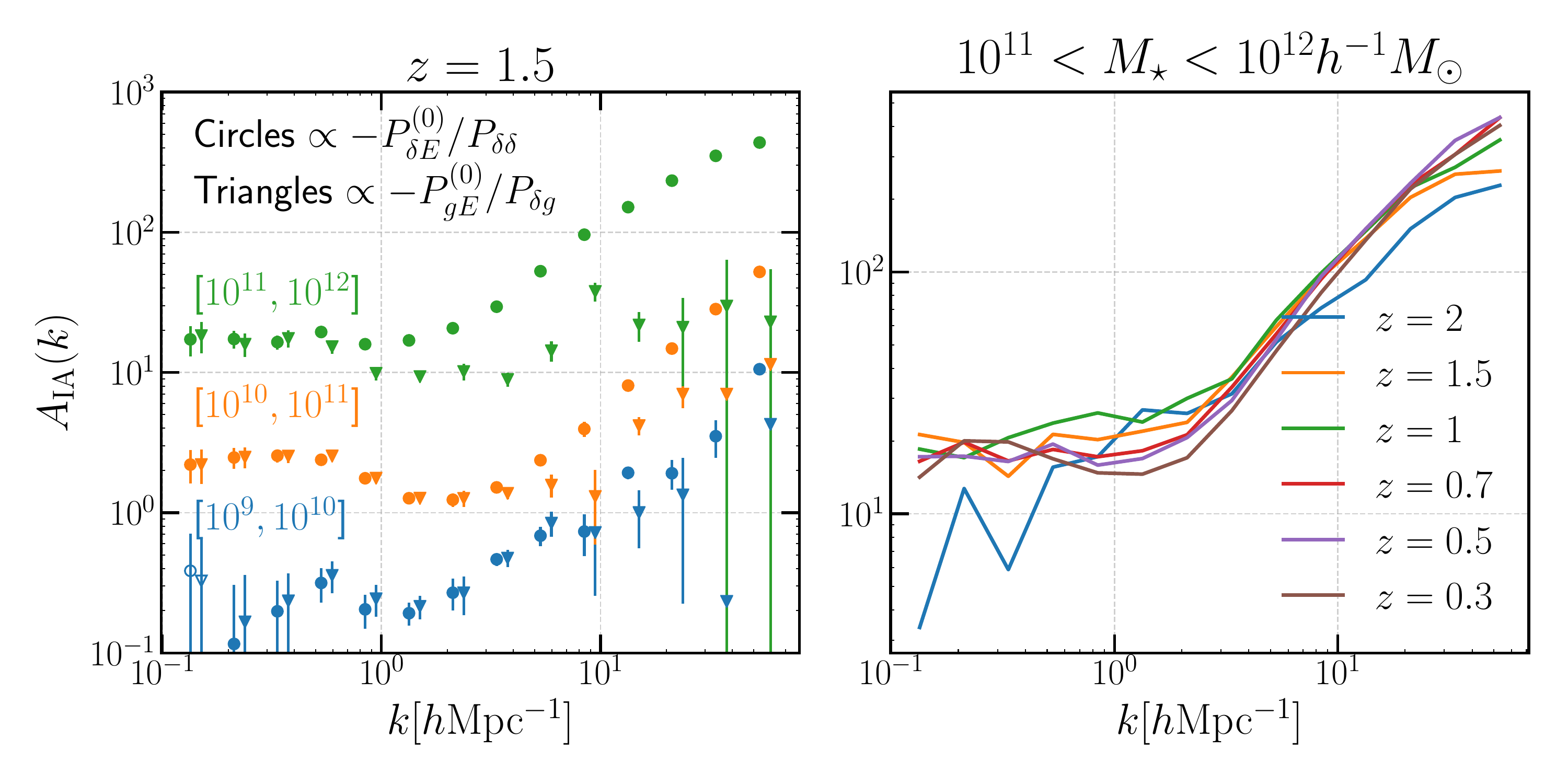}
\caption{{\it Left panel}: The IA factor, defined by 
$A_{\rm IA}(k)\propto -P_{\delta E}^{(0)}/P_{\delta\delta}$ (circles) and $A_{\rm IA}(k)\propto -P_{g E}^{(0)}/P_{\delta g}$ (triangles) according to Eqs.~(\ref{eq:A_IA_dE}) and (\ref{eq:A_IA_gE}) at $z=1.5$, for the same stellar-mass samples as in the previous figure. For illustrative purpose the triangle symbols are slightly shifted in the horizontal direction. The negative values of $A_{\rm IA}(k)$ is shown using the open symbols.
{\it Right}: The redshift dependence of the IA factor for the galaxies in the mass range of $10^{11}<M_\star<10^{12}\msun$ from $z=0.3$ to $z=2$. $A_{\rm IA}$ increases with the stellar mass and does not largely change with the redshifts.}
\label{fig:A_E0_mstarz} 
\end{figure}

In Fig.~\ref{fig:Pk_E0_mstarz} we study how the amplitude of the IA power spectrum varies with different samples of galaxies, especially with different stellar mass samples of galaxies. As can be found from Table~\ref{tab:sample_mstar}, the rms ellipticity of galaxies increases with stellar mass slightly, while it does not vary with redshifts for galaxies of the same mass. The lower panels show the results for the monopole moment of the cross power spectrum of the galaxy shape $E$-mode with the matter density fluctuation field ($\delta$) or the galaxy overdensity field ($\delta_{g}$), i.e., $P_{\delta E}$ or $P_{{ g}E}$, at {three representative} redshifts studied in this work.
We assume the Gaussian covariance to denote the statistical error in each $k$ bin \cite{2004MNRAS.348..897T}. Since we used the folding method for the power spectrum measurements as we stated in Section~\ref{subsec:measurement_method}, we used the actual number of Fourier modes used in the power spectrum measurement to denote the error at each $k$ bin. However, the non-Gaussian errors are not negligible at $k\gtrsim~\mbox{a few }0.1~\hmpci$, so the error bars are meant to just give a guide for the statistical precision of our measurement.

The figure shows that the most massive galaxy sample of $10^{11}<M_\star/\msun<10^{12}$ exhibits the strongest IA signal across the redshift range from $0.3$ to $2$, for both $P_{\delta E}$ or $P_{gE}$. The IA amplitude for the lowest mass 
galaxy sample, $10^{9}<M_\star/\msun<10^{10}$, 
is roughly $2$ orders of magnitude lower than that of the most massive one. This stellar mass dependence of IA power spectrum is consistent with previous results on the halo mass dependence using $N$-body simulations \citep{2020arXiv200412579K} and the results on the luminosity dependence in both observations \citep{Joachimi2011} and hydro simulations \citep{2015MNRAS.448.3522T}.

Since more massive galaxies form at lower redshifts in a hierarchical CDM structure formation scenario, such galaxies tend to preserve the effect of surrounding tidal field more. The difference between $P_{\delta E}$ and $P_{gE}$, for each galaxy sample, is due to the galaxy bias, where the stellar-mass selected galaxies are biased tracers of the underlying mass density field, with the linear bias coefficient $b_{g}>1$ at large scales as shown in the upper panels.

The stellar mass dependence is illustrated in another way in the left panel of Fig.~\ref{fig:A_E0_mstarz}, where we study the IA factor, $A_{\rm IA}(k)$ for galaxies in three different stellar mass ranges at $z=1.5$. 
Here, following Eqs.~(\ref{eq:A_IA_dE}) and~(\ref{eq:A_IA_gE}), we define
$A_{\rm IA}(k)\equiv -[2C_1\rho_{\rm cr0}\Omega_{\rm m}/3D(z)]^{-1} P_{\delta E}(k)/P_{\delta\delta}(k)$ 
or 
$A_{\rm IA}(k)\equiv -[2C_1\rho_{\rm cr0}\Omega_{\rm m}/3D(z)]^{-1} P_{g E}(k)/P_{\delta g}(k)$ 
using the power spectra measured from the simulation. The factor of $2/3$ in $A_{\rm IA}(k)$ comes from $\int_0^1\mathrm{d}\mu~ (1-\mu^2) $.
Note that the $A_{\rm IA}(k)$ factor is free of the galaxy bias parameter, at large scales (small $k$ bins), because $P_{gE}\propto b_1$ and $P_{\delta g} \propto b_1$ on large scales. 
The LA models predicts that $A_{\rm IA}$ is independent of $k$ (scale-independent) in the linear regime, and the NLA model predicts that $A_{\rm IA}$ is scale-independent 
up to the nonlinear scales. The figure shows that $A_{\rm IA}$ appears to be fairly scale-independent at $k\lesssim 0.5~\hmpci$
for the samples, except for the smallest stellar mass sample, implying that the LA or NLA model is valid on these scales. Thus we can estimate the linear IA coefficient, $A_{\rm IA}$, by fitting the $A_{\rm IA}(k)$ values in the three lowest $k$ bins, $k\simeq [0.1,0.3]~\hmpci$, which is given in Table~\ref{tab:sample_mstar}. The linear coefficient $A_{\rm IA}$ increases with stellar mass, showing that the massive galaxies align more with the tidal field.
On the other hand, on the larger $k$ scales, $A_{\rm IA}$ displays scale dependences, indicating a violation of the NLA model. 

Our results might be compared with the previous works \cite{2015MNRAS.450.2195S,Blazeketal2015}, which found that the NLA model works at scales of $2$--$10~h^{-1}{\rm Mpc}$ from the projected IA correlation function measured from the SDSS LOWZ sample, $w_{g+}(r_p)$, which is the integral of $P_{gE}$. Although our results are qualitatively consistent with the previous result, the projected correlation function at a given separation $r_p$ arises from different Fourier modes, and a quantitative comparison of the real- and Fourier-space results is not straightforward.

Interestingly, the left panel shows that the IA coefficients defined from $P_{\delta\delta}$ and $P_{\delta g}$ agree with each other in small $k$ bins in the two-halo term regime. This is encouraging, because the agreement means that the $A_{\rm IA}$ coefficient at small $k$ is free from galaxy bias or more generally small-scale physics involved in galaxy formation. On the other hand, at large $k$ within one-halo term region, $A_{\rm IA} (k)\propto -P_{\delta E}^{(0)}/P_{\delta\delta}$ is higher than $A_{\rm IA} (k)\propto -P_{gE}^{(0)}/P_{\delta g}$, which is also revealed in Ref.~\cite{2015MNRAS.448.3522T} by comparing the projected two-point correlation function between density-shear $\omega_{\delta +}(r_p)$ and galaxy density-shear $\omega_{g+}(r_p)$. This is saying galaxy is more aligned with the matter density field than the galaxy overdensity field, which is also shown by Ref.~\cite{2015MNRAS.448.3522T} using the projected correlation function.

The right panel of Fig.~\ref{fig:A_E0_mstarz} shows that, for the galaxy sample of a fixed stellar mass range, $10^{11}\le M_\star/\msun\le 10^{12}$, the IA factor, $A_{\rm IA}(k)$, has a rather weak redshift evolution. The weak redshift dependence of $A_{\rm IA} (k)$ indicates that the shapes of massive galaxies are imprinted in the matter dominated era or equivalently are determined by the primordial tidal field, as suggested by Ref.~\cite{HirataSeljak2004} \citep[also see][]{2020arXiv200412579K}.
This is consistent with findings in other work. For example, Ref.~\cite{Joachimi2011} found that the IA measured from MegaZ-LRG sample has a redshift dependence of $\propto (1+z)^{-0.3}$, which is rather weak. We checked that the similar weak $z$-dependence is found even if using $P_{gE}$ instead of $P_{\delta E}$. 
The result for $z=2$ looks noisy, and we think that the noisy result is due to the fact that the galaxies are still in the rapid evolving stage, and the galaxy shapes are not yet well established at $z\sim 2$. This may imply that the redshift when IA is imprinted and the galaxy formation redshift might be different. For galaxies of a certain stellar mass, the IA that we observe has an origin of the large scale tidal field which does not evolve much compared to the primordial tidal field. 
That is, the IA effect is imprinted in the matter dominated regime (at a sufficiently high redshift, $z>1$).
However, we think this does not mean that the IA of the galaxy we observe is already there even before the galaxy formed. In fact, Ref.~\cite{2020MNRAS.491.4057B} found the IA signal of massive ellipticals at $z=0$ appeared only after $z=0.5$. The difference between galaxy formation redshift and the IA imprinted redshift is still an open question and more efforts are needed to clarify this point.
In Appendix~\ref{sec:b_K}, we discuss that the linear IA effect can be reconsidered by an alternative IA coefficient, called $b_K$, defined in terms of the density-related tidal field, which has the same dimension as the mass density fluctuation field, rather than the primordial tidal field \cite{Schmidtetal2015}. The $b_K$ definition is, for example, often used in the perturbation theory approach, e.g. Ref.~\cite{Schmidtetal2015,2020arXiv200703670A}. We show the results for $b_K$ for the same galaxy sample as that in the right panel of Fig.~\ref{fig:A_E0_mstarz}, and it displays a greater amplitude at higher redshifts, very much like the linear density bias parameter $b_1$.

\subsection{$M_\star$-limited and SFR-limited samples}
\label{subsec:IA_fixng}

\begin{figure}[!htb]
\centering 
\includegraphics[width=.49\textwidth]{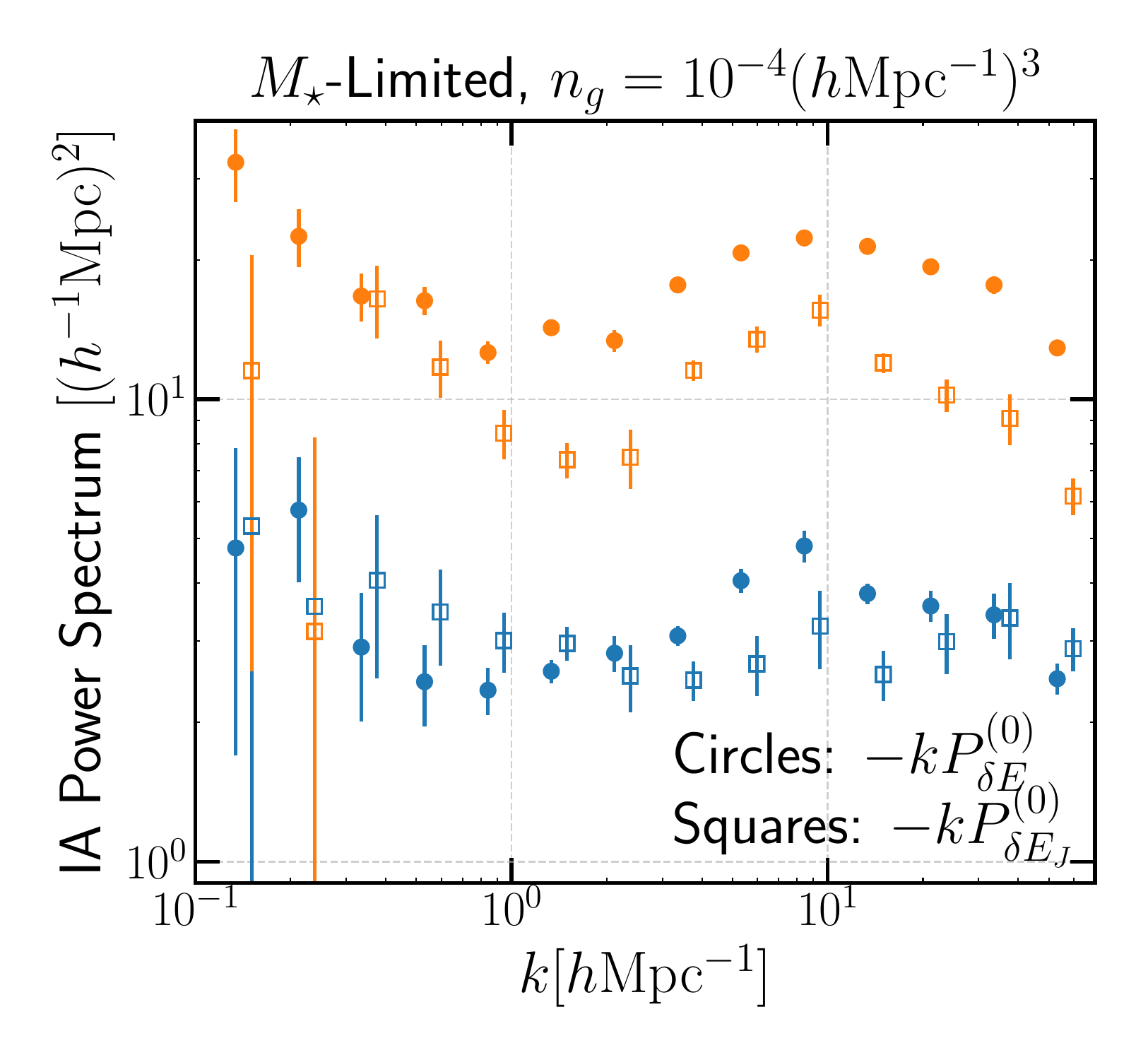}
\includegraphics[width=.49\textwidth]{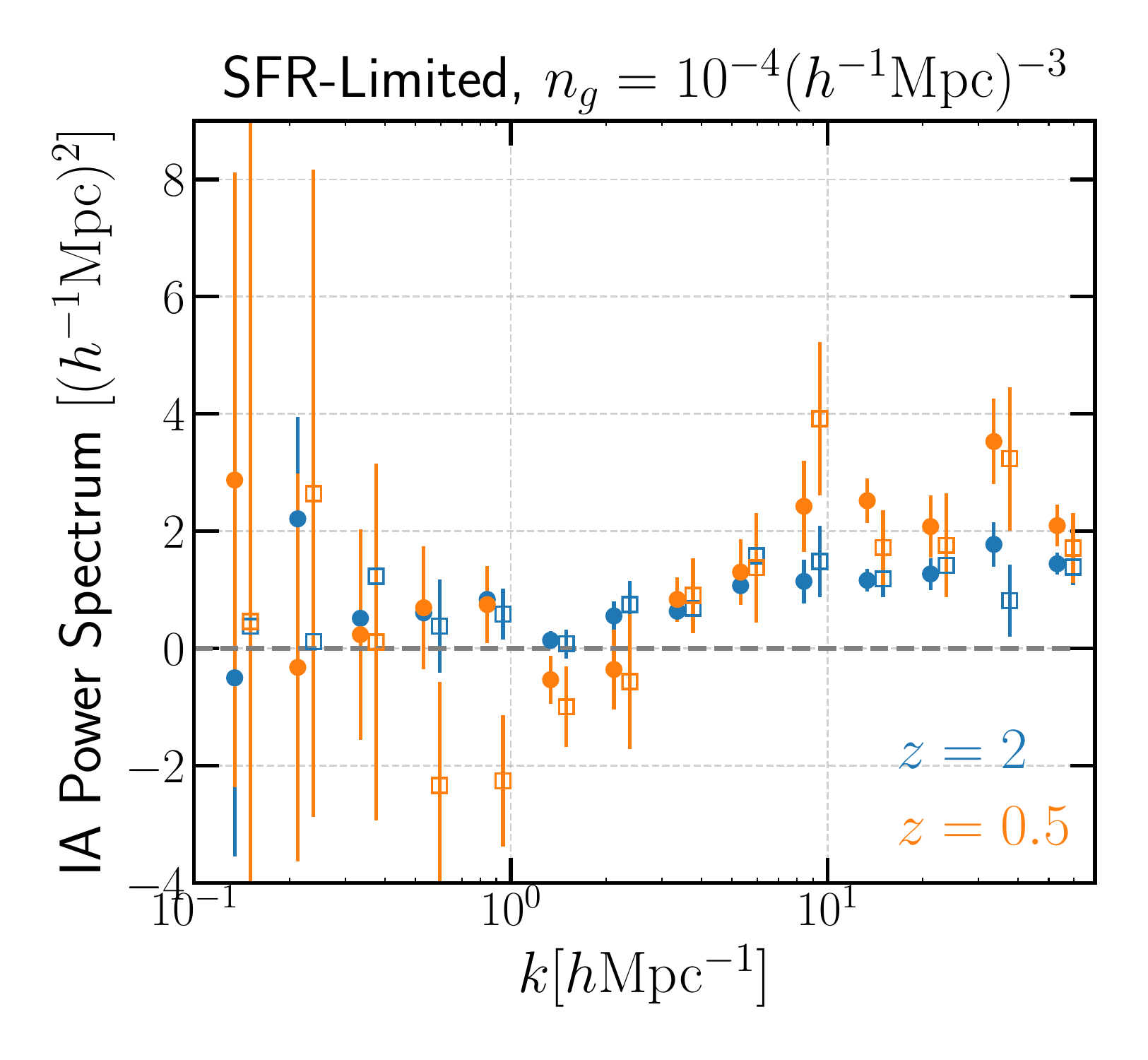}
\caption{The IA cross power spectrum, $P_{\delta E}$, 
for fixed number density samples ($n_g=10^{-4} (\mpc)^{-3}$), ranked by either stellar mass (left panel) or SFR (right panel). Here we show the results for the IA shear calculated by both the reduced inertia tensor ($-kP_{\delta E}$, filled circles) and that of the angular momentum vector ($-kP_{\delta E_J}$, open squares). Note that the $y$ axis of the right panel is not in log scale. The IA alignment for the $M_\star$-limited galaxies characterized by $P_{\delta E}$ and $P_{\delta E_J}$ is clear and strong. There is no IA signal for SFR-limited sample at $k<3\hmpci$.
\label{fig:Pk_E0_fixng}}
\end{figure}
\begin{figure}[!htb]
\centering 
\includegraphics[width=.9\textwidth]{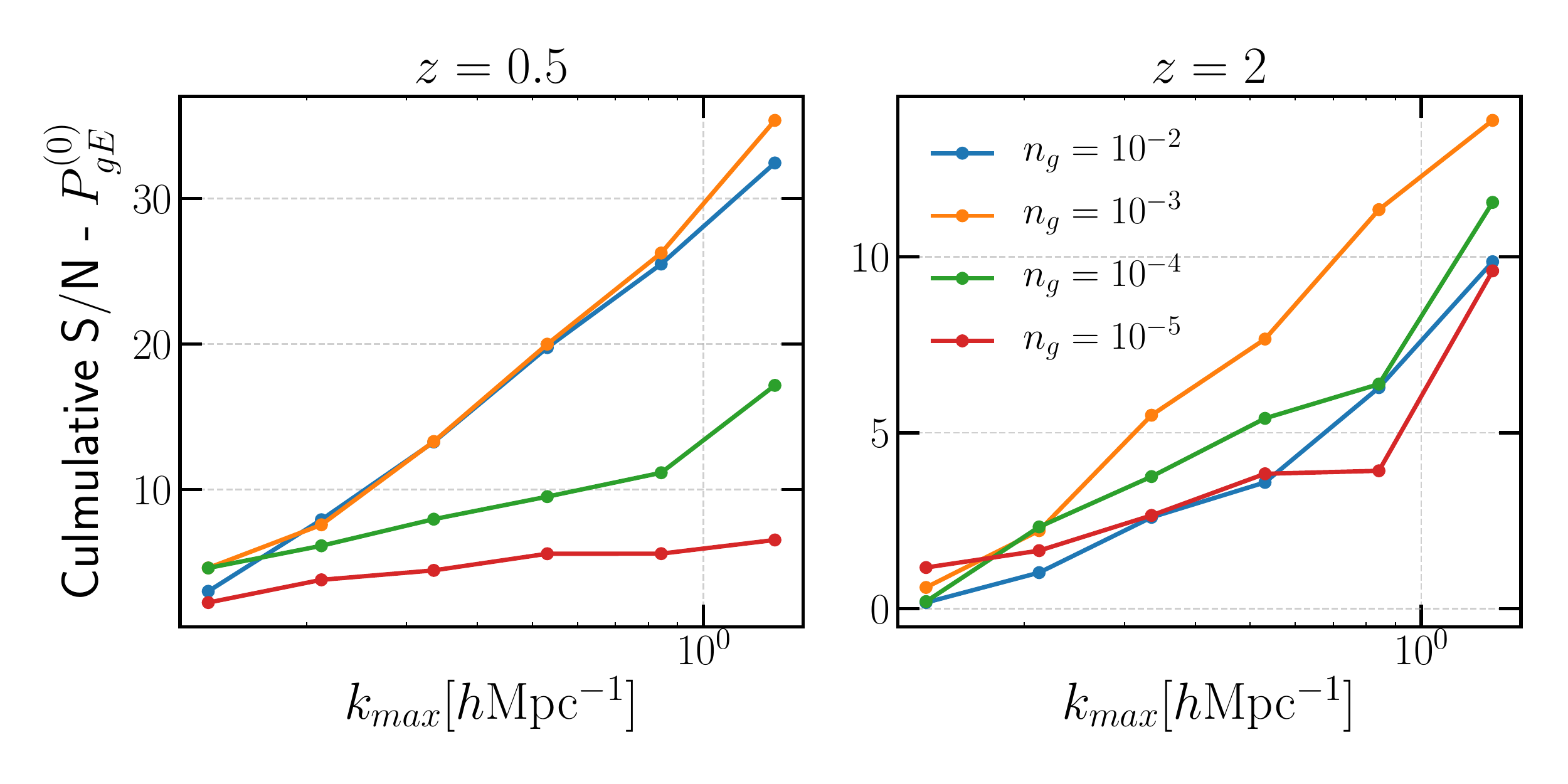}
\caption{The cumulative signal-to-noise ($S/N$) ratio as a function of $k_{\rm max}$ for $M_\star$-limited samples with varying number densities, as indicated by the legend.
The number densities here are in units of $(\mpc)^{-3}$. The cumulative $S/N$ is calculated by integrating the differential $S/N$ in each $k$ bin over $0.1<k/\hmpci<k_{\rm max}$. Here we assume the Gaussian covariance for simplicity. The cumulative $S/N$ ratio is highest when $n_g=10^{-3}(\mpc)^{-3}$.
\label{fig:SN_fixng}}
\end{figure}

Ongoing and planned surveys, such as BOSS \cite{2013AJ....145...10D}, DESI \cite{2016arXiv161100036D} and PFS \cite{2014PASJ...66R...1T}, target either luminous, early-type galaxies or emission-line galaxies. Here an early-type galaxy sample is obtained from a spectroscopic observation of galaxies selected based on color and magnitude cuts \citep[e.g.][]{2001AJ....122.2267E}, and is considered as a proxy of a stellar mass-limited sample, albeit incomplete. On the other hand, an emission-line galaxy (ELG) sample preferentially selects star-forming galaxies \citep[][also see Osato et al. in preparation]{2020arXiv200106560G}.
In this section, we study the IA power spectrum for $M_\star$-limited and SFR-limited samples that resemble samples for actual surveys.

Fig.~\ref{fig:Pk_E0_fixng} shows the IA power spectrum for the two fixed number density samples, 
$n_g\simeq 10^{-4}~(h^{-1}{\rm Mpc})^{-3}$, where galaxies are selected in the ranked list of either $M_\star$ or ${\rm SFR}$ from the top ones.
Here the number density of $10^{-4}~(h^{-1}{\rm Mpc})^{-3}$ is chosen because it roughly corresponds to a target number density of galaxies for ongoing or future surveys.
Note that we calculated the ellipticity using either the reduced inertia tensor of stellar distribution or the stellar angular momentum vector, as defined in Eqs.~(\ref{eq:e_inertia}) and ~(\ref{eq:e_J}). We use $P_{\delta E}$ and $P_{\delta E_J}$ to denote the resulting IA power spectra separately. 

For the $M_\star$-limited sample, there exists a clear IA signal at all redshifts for both $P_{\delta E}$ and $P_{\delta E_J}$. The IA strength for $P_{\delta E}$ at $z=0.5$ and $2$ are $A_{\rm IA}=24.05\pm 2.01$ and $9.69\pm2.02$, respectively. The stronger IA at $z=0.5$ is mainly driven by the existence of the more massive galaxies at low redshift given a fixed number density. The mean stellar mass of the $M_\star$-limited sample 
with $n_g=10^{-4}(\mpc)^{-3}$ is $10^{11.39}\msun$ and $10^{11.04}\msun$ at $z=0.5$ and $2$, respectively.
As discussed in the above subsection, massive galaxies tend to be more aligned with the surrounding tidal field.
The IA power spectrum computed from the angular momentum, for the stellar mass limited samples, clearly shows a non-zero signal in small $k$ bins. The non-zero IA signal at these large scales seems to obligate the prediction of the quadratic alignment model assuming linear structure growth (see Section~\ref{subsec:QA}), because the quadratic alignment model assuming linear structure growth predicts $P_{\delta E_J}/P_{\delta\delta} \rightarrow 0$ at the limit of $k\rightarrow 0$. It indicates that the angular momentum of galaxies has a physical correlation with the large-scale tidal field via the nonlinear gravitational interaction.
If the angular momentum vector is perfectly aligned with the minor axis of the spheroid of the inertia tensor and the stellar distribution is disk-like, we would expect $P_{\delta E_J}$ to be the same as $P_{\delta E}$. However, there exists a misalignment between the angular momentum vector and the minor axis of the galaxy, although the alignment signal is strong (see Fig.~2 in Ref~\cite{2015MNRAS.452.3369C}). And the galaxy usually does not have a perfect disk. Thus $P_{\delta E_J}$ can be different from $P_{\delta E}$, but they show a similar $k$-dependence.
The generally good agreement on the amplitude and $k$-dependence between $P_{\delta E_J}$ and $P_{\delta E}$, indicates a rather good alignment between the angular momentum vector and the minor axis of the inertia tensor.
Note the central galaxy fraction of the $M_\star$-limited samples is $\sim 90\%$ across $z=0.3$ to $z=2$. This is further supported by results shown in the lower left panel of Fig.~\ref{fig:Pk_E0_satcen}, where the agreement between $P_{\delta E}$ and $P_{\delta E_J}$ for massive centrals are surprisingly good.

Fig.~\ref{fig:SN_fixng} gives the corresponding cumulative signal-to-noise ($S/N$) ratio of $P_{gE}$ for the $M_\star$-limited samples of $n_g=10^{-4}~(h^{-1}{\rm Mpc})^{-3}$.
The cumulative $S/N$ ratio expected for a measurement of $P_{gE}$ is calculated by summing up the $S/N$ ratios in each $k$ bin over a range of $k_{\rm min}<k<k_{\rm max}$, defined by
\begin{equation}
    \left(\frac{S}{N}\right)^2 \equiv  \sum_{k_i = k_{\rm min}}^{k_{\rm max}} \bar{P}^{(0)}_{gE}(k_i) 
    \left[{\bf C}\right]^{-1}_{ij} \bar{P}^{(0)}_{gE}(k_j),
    \label{eq:snr_ssc}
\end{equation}
where ${\bf C}$ is the covariance matrix between the monopole moments of of power spectra and $[{\bf C}]^{-1}$ is the inverse of the covariance matrix, and we adopt $k_{\rm min}=0.1~\hmpci$ throughout this paper. Here we assume a Gaussin covariance matrix \cite{2013PhRvD..87l3504T}. The cumulative $S/N$ in TNG300 volume for these samples can reach 30 at $z=0.5$ for $k_{\rm max}=1\hmpci$. This corresponds to $S/N\simeq 320$ for a cosmological volume of $1~(h^{-1}{\rm Gpc})^3$ because $S/N$ scales with a volume as $S/N\propto V^{1/2}$ for the Gaussian covariance case. Note that the TNG300 has a volume of $(205~h^{-1}{\rm Mpc})^3$. We will again discuss the $S/N$ ratio in Section~\ref{sec:survey}.  

For the SFR-limited sample, the IA at $k<2\hmpci$ is rather weak and noisy. The IA strength of $P_{\delta E}$ at $z=2$ and $z=0.5$ is $A_{\rm IA}=0.54\pm 0.53$ and $A_{\rm IA}=0.47\pm 1.85$, indicating that the orientation of the ELGs are pointed randomly with respect to the large-scale structure at $0.3<z<2$. This is qualitatively consistent with the IA observations using blue galaxies at low ($z<0.2$) and intermediate redshifts ($z\sim 0.6$) \citep{2007MNRAS.381.1197H,2011MNRAS.410..844M}, since most star-forming galaxies are intrinsically blue as well. Our results show a null detection at even higher redshift. However, a quantitative comparison with observation requires a careful treatment of the dust, $K$-correction and other observation effects, which is beyond the scope of this work. At $k>2\mpc$, the IA exists for both $P_{\delta E}$ and $P_{\delta E_J}$, indicating that both the shape and angular momentum of those ELGs align with their local tidal field. The mean stellar mass of SFR-limited samples are $\sim 10^{10.4}\msun$, and the satellite fraction ranges from $20\%$ to $37\%$ for SFR-limited samples at $z=2$ to $z=0.3$. The small-scale IA signal are thus a combination of the contribution from centrals and satellites. For centrals, the small-scale tidal field is dominated by the host dark matter halo. Previous work shows that both the shape and angular momentum of central galaxies tend to align with their host halos \cite{2015MNRAS.453..721V,2014MNRAS.441..470T,2017MNRAS.472.1163C}, although the misalignment does exist and can be large in some cases. The alignment strength is found to be stronger with the inner halo region. The small-scale tidal field of satellites, on the other hand, is dominated by the central galaxies. Previous work using both $N$-body and hydrodynamic simulations \cite{2008ApJ...675..146F,2010MNRAS.405.1119K,2015MNRAS.453..469T} found that they have a preferred orientation towards centrals in the same host halo. Such radial alignment has also been detected by Ref.~\cite{2018MNRAS.474.4772H} using satellites in redMaPPer clusters, however the detection depends on the shape measurement method. We will come back the IA of centrals and satellites in Section \ref{subsec:IA_censat}.

\subsection{Morphology dependence}
\label{subsec:IA_morph}
\begin{figure}[!htb]
\centering 
\includegraphics[width=.9\textwidth]{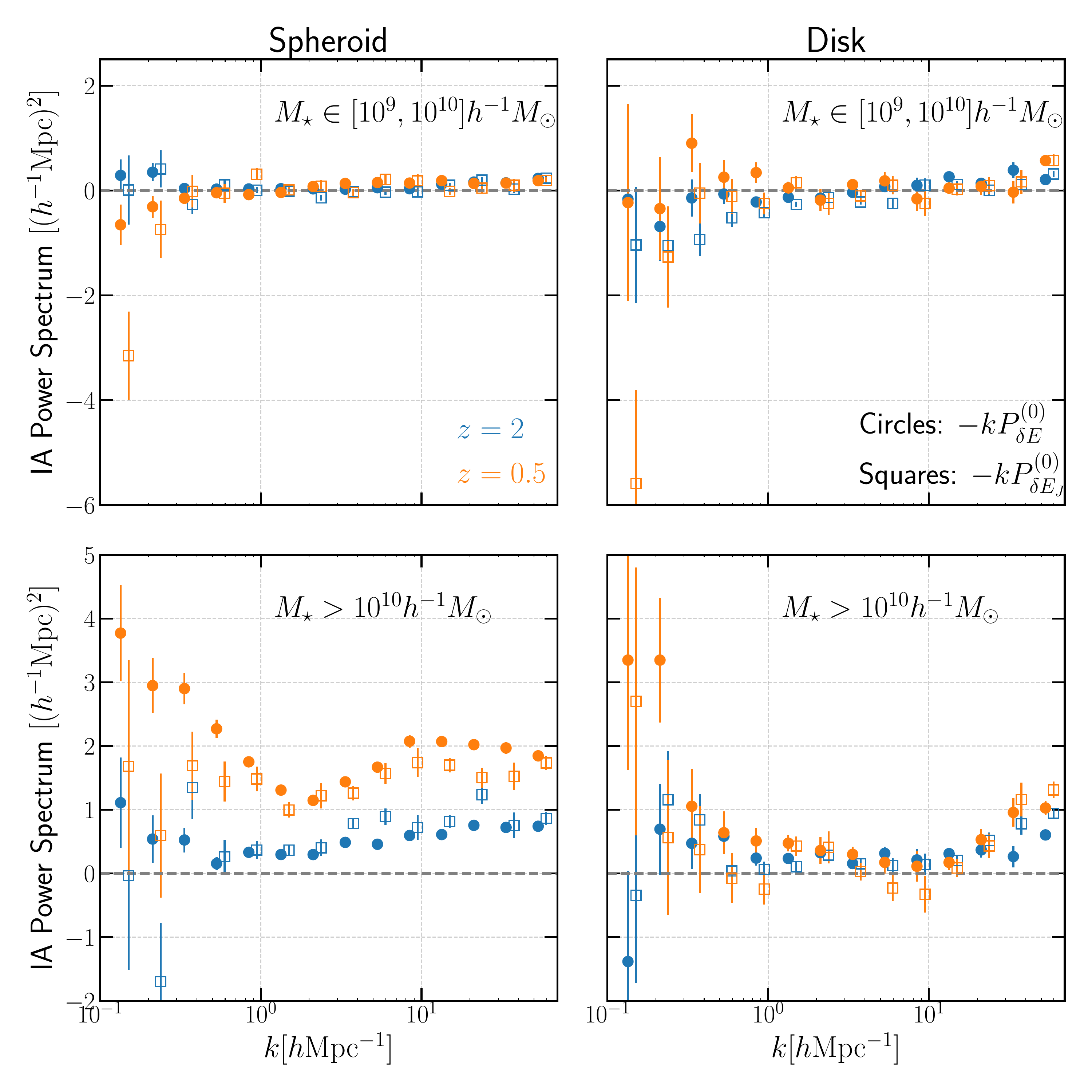}
\caption{Similar to Fig.~\ref{fig:Pk_E0_fixng}, but this figure shows the IA power spectra for the galaxy samples of different morphologies; spheroids ($\kappa_{\rm rot}<0.45$) and disks ($\kappa_{\rm rot}>0.55$) at $z=2$ (blue symbols), and $z=0.5$ (orange). The upper panels are for galaxies with $M_\star>10^9\msun$, and the lower panels are for galaxies with $M_\star>10^{10}\msun$. Filled circles show the cross spectra between the matter density and the IA shear calculated by the reduced inertia tensor, $-kP_{\delta E}$, while open squares show the cross spectra between the mass density field and the IA shear calculated by the angular momentum, $-kP_{\delta E_J}$. The horizontal gray dashed line denotes a zero signal of the IA power spectrum.
\label{fig:Pk_E0_morph} }
\end{figure}

The orientations of galaxies of different morphology are thought to originate from different mechanisms. 
In this subsection we study the IA power spectra for the spheroid- and disk-like galaxies. The sample properties and the measurement results are summarized in Table~\ref{tab:sample_disk_spheroid}. Disks tend to have higher $e_{\rm rms}$ compared to spheroids, and there are more star-forming activities in disks compared to spheroids. Most massive spheroids at $z<1$ are quiescent.

Fig.~\ref{fig:Pk_E0_morph} shows the IA power spectra for spheroid- and disk-like galaxies at $z=2$ and $z=0.5$. First of all, the massive spheroid-like galaxies display a significant signal of $P_{\delta E}$, indicating a strong alignment between the major axis of the projected ellipse and that of the tidal field across all the scales we study, i.e., the range of $0.1<k<60\hmpci$, from $z=2$ to $z=0.3$. While the low-mass spheroidal galaxies show much weaker alignment, and the alignment is even reversed a little bit from high redshift to low redshift, as indicated by the negative $A_{\rm IA}$ in Table~\ref{tab:sample_disk_spheroid}.
In contrast, the IA signal of $P_{\delta E}$ for low-mass disks does not exist, while it does exist for massive disks, although with large uncertainties as shown in Table~\ref{tab:sample_disk_spheroid}. The alignment signal is weaker for disks at given stellar mass ranges.

Our predictions for IA of disk galaxies in TNG seems to be inconsistent with the results in Ref.~\cite{2016MNRAS.462.2668T}, where they did not find any significant signal in the projected two-point correlations of IA at scales above $0.1~\mpc$ in MassiveBlack-II and Illustris cosmological hydrodynamical simulations for both low-mass and massive disks. However, there are two points that is worth clarifying. First, the division of the galaxy morphology into disk and spheroid is different. We use the $\kappa_{\rm rot}=\frac{K_{\rm rot}}{K}$ to distinguish the rotation dominated, disk-like galaxies from the dispersion dominated spheroid-like galaxies, while Ref.~\cite{2016MNRAS.462.2668T} uses the bulge to disk ratio $B/T$. These two division methods do not have one-to-one correspondence. Second, the astrophysical model employed in those simulations are quite different, which might be responsible for the different IA signals of disks among those simulations. Interestingly, our results seem to agree with the observed weak IA of the luminous blue galaxies in Ref.~\cite{2007MNRAS.381.1197H}. However, a quantitative comparison with the observations is beyond the scope of this work.

The cross correlations between the matter field and the shear field calculated by the angular momentum vector, $P_{\delta E_J}$, show a more complicated trend. As discussed in Section \ref{subsubsec:e_J}, $P_{\delta E_J}$ is an indicator of the spin alignment with tidal field across the scales we measured at the level of two-point correlation statistics. For low mass galaxies ($M_\star\in [10^{9},10^{10}]\msun$, both spheroids and disks), the sign of $P_{\delta E_J}$ is reversed at $k<3~\hmpci$, as shown by the open squares in the upper panels of Fig.~\ref{fig:Pk_E0_morph}. Such inverted IA signal, characterized by a negative $A_{\rm E_J}$, is stronger for low-mass disks than low-mass spheroids, especially at high redshift, as shown in Table~\ref{tab:sample_disk_spheroid}. Massive disks and spheroids, on the other hand, show a positive $A_{\rm E_J}$. 
Our result indicates that the spin vector of the low mass galaxies tend to align with filament, while the spin vector of the massive spheroid is perpendicular to the filament.
The inversed signal of $P_{\delta E_J}$ for low mass disks and spheroids is consistent with the expectation of the so-called spin-flip phenomenon. Numerical simulations revealed that low mass galaxies or halos tend to have their spin aligned with the filament, while the spin of high mass galaxies or halos is flipped and become perpendicular to the filament \cite{2018MNRAS.481.4753C,2015MNRAS.452.3369C,2015ApJ...813....6K}.
The transition mass is estimated to be $5\times 10^{12}\msun$ for halos \cite{2012MNRAS.427.3320C} and $10^{10.5}\msun$ for galaxies \citep[][based on Horizon AGN simulation]{2014MNRAS.444.1453D}
at $z=0$. The flip of the spin is found to be explained by the varying halo accretion history that depends on the halo mass and environment \cite{2015ApJ...813....6K,2017MNRAS.468L.123W,2018MNRAS.473.1562W}. For example, mergers play a specific role in flipping the spin of massive objects to be perpendicular to the filament \cite{2014MNRAS.445L..46W,2012MNRAS.420.3303B,2012MNRAS.420.3324B}, since the late-time accretion happens mainly along the filament. Lots of efforts have also been made in confirming the spin-flip phenomenon in observation \cite{2007ApJ...670L...1L,2013MNRAS.428.1827T,2013ApJ...775L..42T,2017ApJ...837...31K,2020MNRAS.491.2864W}. Note that, roughly 25\% of the low mass galaxies are disks (see Table~\ref{tab:sample_disk_spheroid}), and they show a clear signature of the flipped spin alignment across the redshift ranges that we studied in this work. Low mass spheroids, on the other hand, only have detection at $z=0.3$ and $z=0.5$. Thus the flipped spin alignment of low-mass galaxies is dominated by the disks. On the other hand, for massive galaxies at $z>1$, the number density and $A_{\rm E_J}$ of disks are larger or comparable than that of the spheroids. At $z<1$, there are more massive spheroids and their $A_{\rm E_J}$ is larger. In brief summary, the low mass spin-alignment is dominated by disks, especially at high redshift, while the high mass spin-alignment signal is dominated by disks at $z>1$, but dominated by spheroids at $z<1$.
However, we should point out that $A_{E_J}$ is calculated using Eq.~(\ref{eq:A_IA_dE}), which is based on the LA for the alignment of shape, not angular momentum. 

Recently, Ref.~\cite{2020arXiv200910735S} studied the IA of galaxies from IllustrisTNG, MassiveBlack-II, and Illustris-1 hydro-simulations. They constrained the model parameters of NLA and TATT (tidal alignment + tidal torque) model parameters for a variety of galaxy samples at $0<z<1$ by fitting the projected 2-point correlation function. They found no significant evidence for non-zero values of the tidal torquing amplitude, $A_2$, in IllustrisTNG. At a first glance, our result seems to disagree with each other. However, the ellipticity is calculated by using the projected shape of stellar distribution, which is different from the spin based ellipticity that is discussed here. Thus, even if the tidal torquing amplitude $A_2$ in TATT is zero, it does not necessarily mean that the spin-matter field correlation is zero. 
Shortly after that, Ref.~\cite{2020arXiv201007951Z} tested the tidal torquing mechanism by fitting directly the correlation between ellipticity based on angular momentum and tidal field (smoothed at ~$1\mpc$) at $z=0$ and $z=1$ for spirals. Both of the results showed an alignment between spin and surrounding tidal field for massive spirals at $z=1$. However, the reverted spin-alignment for low mass spirals seen in our work is not shown in Ref.~\cite{2020arXiv201007951Z}. We think this might be caused by the different ways to measure the IA alignment, and we intend to leave the effects of varying measuring methods to future works.

\subsection{Central versus satellite}
\label{subsec:IA_censat}

\begin{figure}[!htb]
\centering 
\includegraphics[width=.9\textwidth]{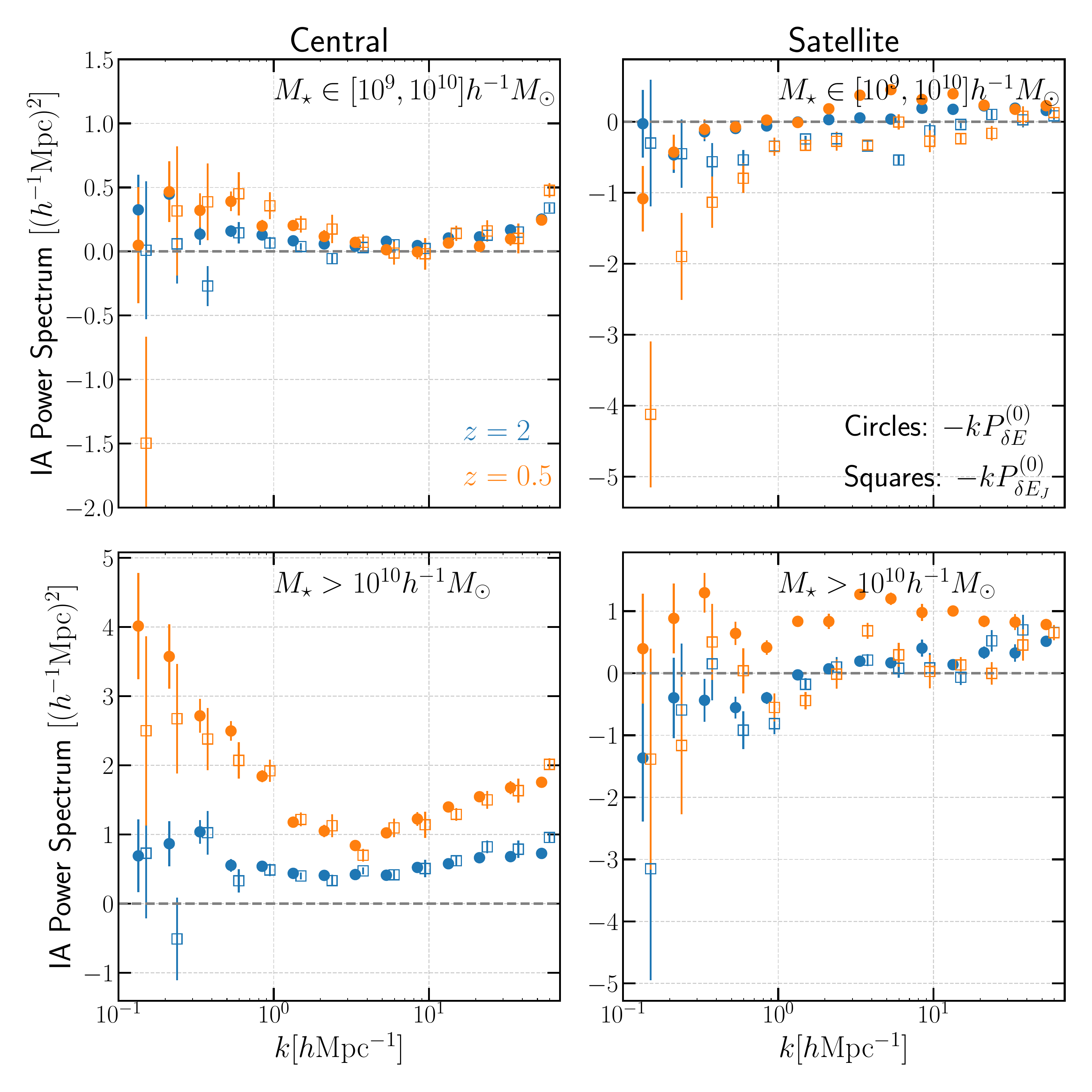}
\caption{Similar to the previous figure, but the IA power spectrum for central and satellite galaxies.
\label{fig:Pk_E0_satcen} }
\end{figure}
We now study the IA power spectra for the central and satellite galaxies. The sample properties and results are summarized in Table~\ref{tab:sample_central}.
In Fig.~\ref{fig:Pk_E0_satcen}, we show the IA power spectra, $P_{\delta E}$ and $P_{\delta E_J}$, for the central galaxies and satellite galaxies. The alignment between the major axis of centrals and tidal field is rather clear up to $z=2$, and its alignment strength increases significantly with the stellar mass. The alignment between spin and tidal field of massive centrals, indicated by $P_{\delta E_J}$, follows closely with $P_{\delta E}$, i.e. the spin vector of massive centrals is perpendicular to the direction of filament. The good agreement between those two suggests that the angular momentum vector of these centrals are very well aligned with the minor axis of the inertia tensor spheroid. However, no spin-matter field alignment is detected for low mass centrals. 
The small scale alignment of centrals is caused by the alignment between central galaxy and its host halo. By inspecting the $A_{\rm IA} (k)$ for centrals, we found the alignment strength increases with increasing $k$, i.e. the central aligns with the central region of host halos more strongly. However, Ref.~\cite{2018MNRAS.481.4753C} found the spin of low-mass ($M_\star<10^{10.3}\msun$) central tends to be parallel with the filament, while the massive central tends to be perpendicular with filament using Horizon-AGN. Such flipped spin-alignment of centrals is not seen in our results. This might be caused by the quite different galaxy formation physics employed in those two simulations.

Interestingly, we see an inverted IA signal of both $P_{\delta E}$ and $P_{\delta E_J}$ for the low-mass satellites, although the detection is weak $\sim 2\sigma$ (see Table~\ref{tab:sample_central}). The massive satellite does have a positive $A_{\rm IA}$ at $z<2$, but with large error bars. Massive satellites have null signal of the spin alignment.
The alignment within one-halo region indicated by $P_{\delta E}$ is clear across all redshift and stellar mass ranges. Since the small scale tidal field is dominated by the central galaxies of host halo, the small scale alignment indicates that the major axis of satellites tend to point towards their centrals. Such a radial alignment is also found to be dependent on the radial distance to the centrals, with satellites in the inner region showing the strongest alignment. This radial dependent alignment of satellites at small scales is also seen in other work and observation \cite{2015MNRAS.453..469T,2020arXiv200105962T,2015MNRAS.450.2195S}. 

However, the IA of satellites at large scales do not reach clear consensus. Ref~\cite{2015MNRAS.453..469T} found a non-vanishing IA alignment signal, characterized by the projected two point correlation function between the density field (traced either by matter or galaxies) and galaxy shape, for satellites at large scales in MassiveBlack-II simulation. Ref~\cite{2018MNRAS.481.4753C} showed that at high $z$, the spin of satellites tend to align with the filament, which is in good agreement with what we found for low-mass satellites. The spherical halo model predicts a null satellite alignment signal on large scales  \cite{2013MNRAS.431..477J,2010MNRAS.402.2127S}, but in the spheroidal halo model, such large scale alignment of satellites exists \cite{2020arXiv200302700F}.
The diverged prediction among various models on the satellite/central IA implies that it can be used as a promising tool to constrain the physics by comparing to the forthcoming observations.

\section{Implication for Ongoing/Future Surveys}
\label{sec:survey}
\begin{figure}[!htb]
\centering 
\includegraphics[width=.95\textwidth]{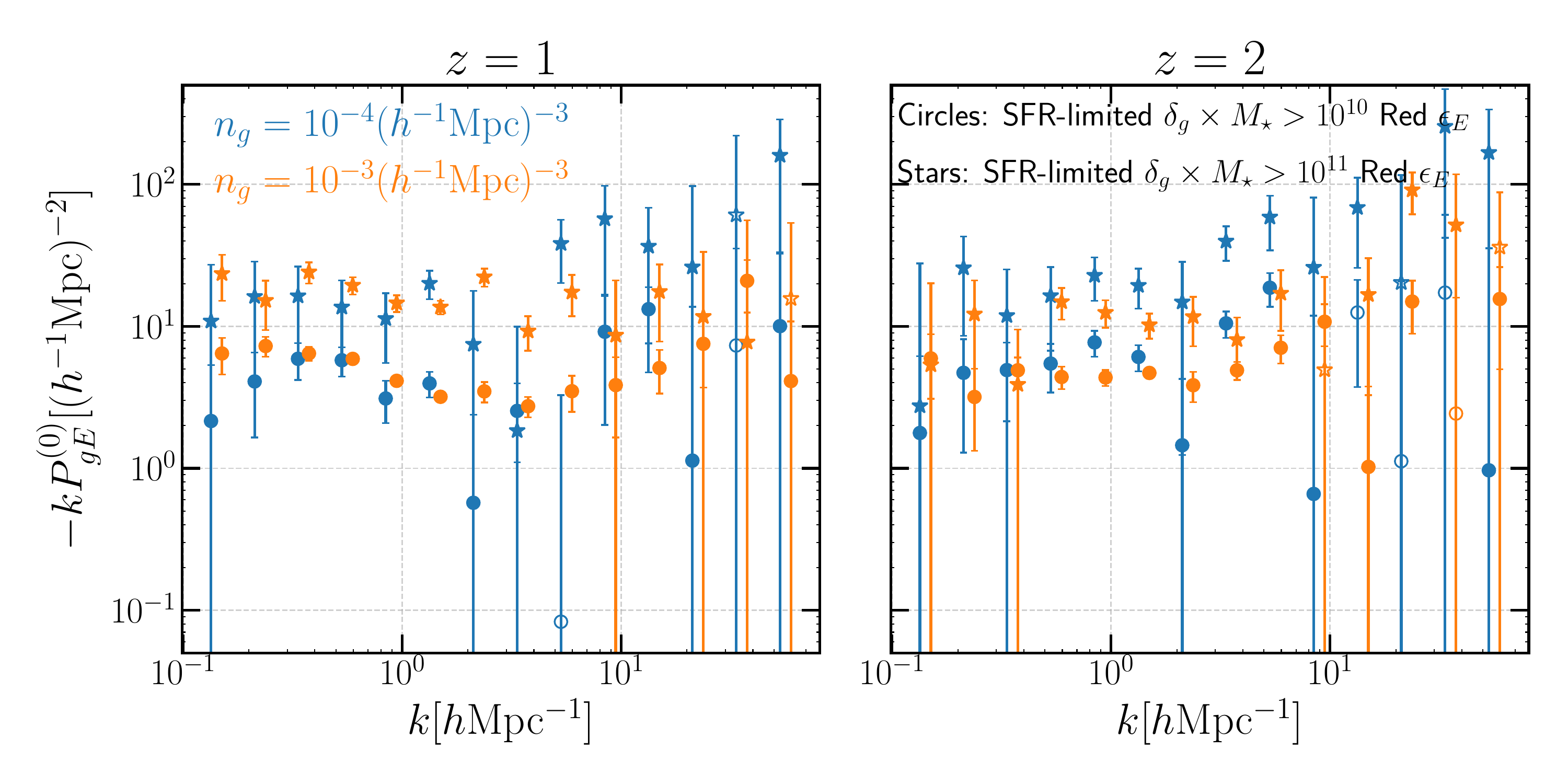}
\caption{\label{fig:Pk_E0_gE} The IA cross power spectrum, $-kP_{g E}$, expected when we use massive, stellar-mass limited galaxies 
for the shape tracers ($\epsilon_E$) and use the fixed-number density sample of SFR-ranked galaxies for the density tracers ($\delta_g$). The circle and star symbols are the results for the stellar-mass limited sample with $M_\star>10^{10}\msun$ or $M_\star>10^{11}\msun$, respectively, and the respective orange and blue symbols are the results for the SFR samples with $n_g=10^{-3} (\mpc)^{-3}$ or $10^{-4} (\mpc)^{-3}$. The open symbols represent the negative values of $-kP_{g E}$.}
\end{figure}
\begin{figure}[!htb]
\centering 
\includegraphics[width=.9\textwidth]{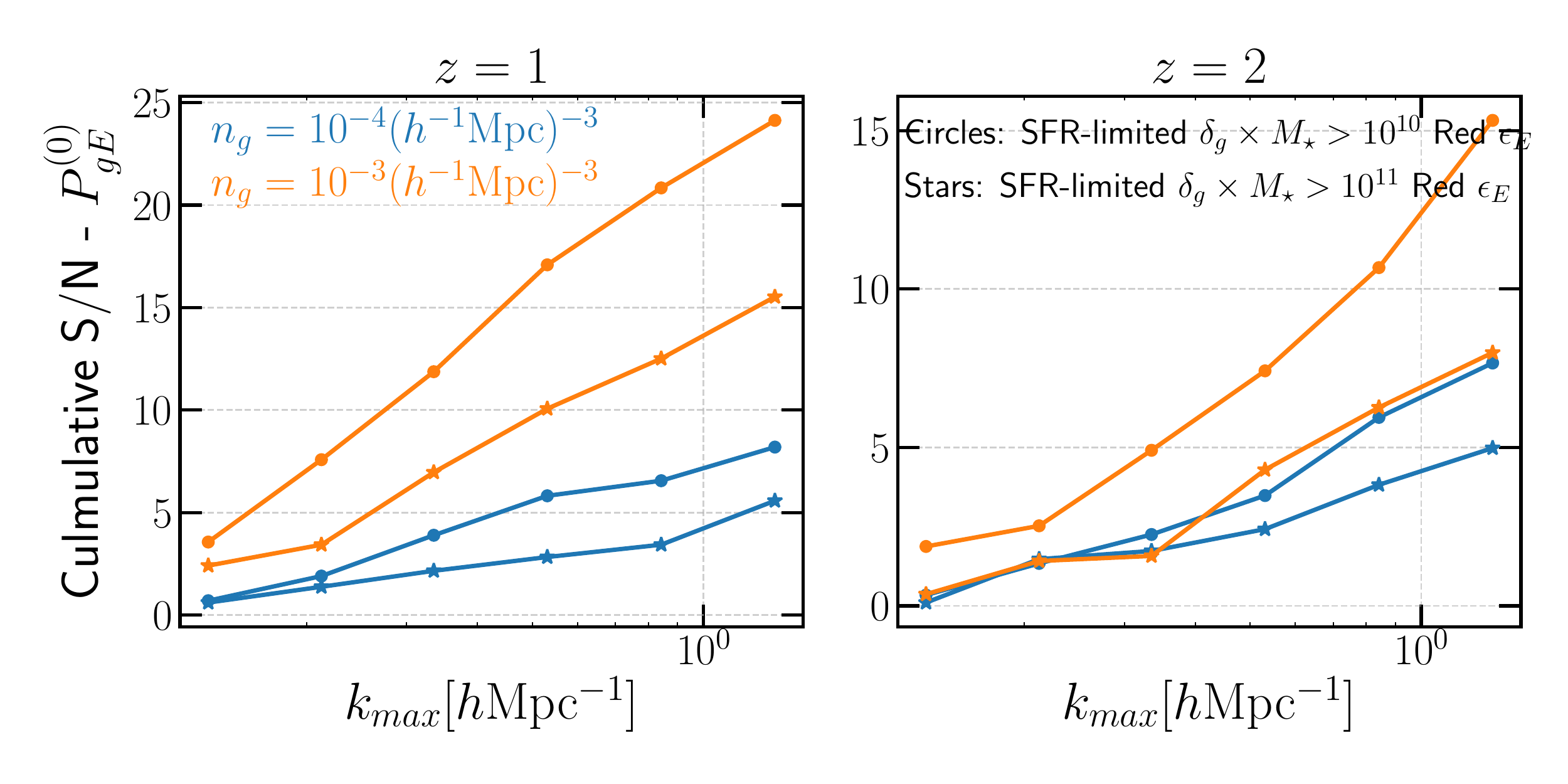}
\caption{\label{fig:SN_brm} The cumulative S/N ratio of the IA power spectra shown in previous figure.}
\end{figure}

Ongoing/future wide-area galaxy surveys will provide a large dataset of spectroscopic galaxies, enabling us to probe the large-scale structure at high redshift to an unprecedented statistical precision. For example, the cosmology program of PFS survey will map about $4$ million [OII] emitting galaxies at $0.6\leq z\leq 2.4$ over a $1400$ ${\rm deg^2}$ region. 
On the other hand, the galaxy evolution program of PFS survey will make a spectroscopic observation for galaxies over $0.7\lesssim z\lesssim 2$, down to much fainter magnitudes with longer exposure, over about $15$ ${\rm deg^2}$.
The PFS survey region is fully covered by the Subaru HSC imaging survey, which possess a superb image quality allowing for accurate measurements of individual galaxy shapes.  
In addition, the HSC and PFS survey regions have an overlap with the SDSS/BOSS spectroscopic survey and the DESI survey, which include a sample of luminous red galaxies up to $z\simeq 1.2$.  
Thus the Subaru HSC and PFS surveys, further in combination with the BOSS/DESI data, will enable us to study the IA correlations for various types of galaxies, e.g., ELGs, stellar mass-limited samples, and various different subsamples 
divided based on their properties (stellar mass, color, etc.).

Fig.~\ref{fig:SN_fixng} shows the cumulative $S/N$ of the IA power spectrum as a function of $k_{\rm max}$ for the $M_\star$-limited samples of different number densities, where we select galaxies from the ranked list of $M_\star$ in the TNG300 simulation
until the number density matches, $n_g=10^{-2}$, $10^{-3}$, $10^{-4}$, or $10^{-5}{(\mpc)^{-3}}$, respectively. 
Here we consider the $S/N$ value for the cross power spectrum $P^{(0)}_{gE}$ as it is a direct observable from the survey regions where both the spectroscopic and imaging data are available. The cumulative $S/N$ can reach $8$ and $15$ for the $M_\star$-limited sample of $n_g=10^{-4}~(h^{-1}{\rm Mpc})^{-3}$ at $z=2$ and $z=0.5$ in the TNG300 volume for $k_{\rm max}=1~\hmpci$. The $S/N$ value does not increase {monotonically} with $n_g$. 
The sample of $10^{-3}~(h^{-1}{\rm Mpc})^{-3}$ has a highest $S/N$ value, and this is encouraging as some of future surveys can reach such a number density for early-type galaxies. 
However, we should mention that we assume a Gaussian covariance matrix while calculating the $S/N$ ratio, and this overestimates the $S/N$ value by up to $50\%$ at $k_{\rm max}\gtrsim \mbox{a few}~0.1~\hmpci$ (See Section~$4.4$ in Ref.~\cite{2020arXiv200412579K} for more details).

For a measurement of the IA power spectrum $P_{gE}$, we do not necessarily use the same sample of galaxies. As we have shown, early-type, massive galaxies have strongest IA correlations. On the other hand, ELGs make it relatively easy to achieve a higher number density, especially for high redshifts. Hence, for the IA measurement, we could use early-type galaxies for shape tracers, while use ELGs for density tracers in large-scale structures. Here we study this case that we use different types of galaxies for the IA measurement. Fig.~\ref{fig:Pk_E0_gE} shows the IA power spectra where we use the massive red galaxies, in analogy to the LRGs, as shape tracers and use SFR-limited sample as density tracers. The cumulative $S/N$ ratio of these samples in TNG300 volume are shown in Fig.~\ref{fig:SN_brm}.
The IA exists for all the samples, and its amplitude is stronger when more massive red galaxies are used as tracers of the ellipticity field. However, the $S/N$ ratio decreases (as shown in Fig.~\ref{fig:SN_brm}) when more massive red galaxies are used as the density tracers too, because their number density is too low. The IA power spectrum strength seems to be independent of the choice of the tracer for the density field. 
For a cosmological volume covering $1~(h^{-1}{\rm Gpc})^3$ or greater, the expected $S/N$ is larger than what is shown 
in Fig.~\ref{fig:SN_brm} by more than a factor of 11. Hence the future surveys can achieve a significant detection of the IA power spectrum against different samples of galaxies.

\section{Discussion and Conclusion}
\label{sec:con}
We have studied the shape and spin IA characterized by the 3D power spectra for various galaxy samples in TNG300. Our main findings are as follows.

\begin{itemize}
    \item The IA power spectrum of galaxies has greater amplitudes with more massive stellar-mass galaxy samples.
    The linear-scale amplitude coefficient $A_{\rm IA}$, 
    defined in terms of the primordial tidal field,
    yet does not evolve strongly with the redshift for galaxies of the same stellar mass. 
    
    \item At small scales, the IA deviates from the expectation of the NLA/LA models. The IA alignment factor, defined as
    $A_{\rm IA}(k)\propto P_{\delta E}/P_\delta$ or $P_{gE}/P_{\delta g}$, has greater amplitudes with increasing $k$.
    
    \item The IA for galaxies of number density $n_g=10^{-4}(\mpc)^{-3}$ ranked by $M_\star$ is strong from $z=0.3$ to $z=2$, with a promising cumulative $S/N$ ratio in the survey volume covered by future surveys. However, no detection of IA for ELGs ($n_g=10^{-4}(\mpc)^{-3}$ ranked by SFR) is found. 
    
    \item The cross power spectrum between the density field and the elliptictiy measured using the angular momentum (spin) vector of galaxies, $P_{\delta E_J}$, is non-zero at $k<1h{\rm Mpc^{-1}}$. This is in disagreement with the prediction of linear quadratic-alignment model, indicating that some degree of the angular momentum is triggered by the nonlinear evolution of matter density field. 
    
    \item The shape and spin IA effects depend on the galaxy morphology. Massive disks and spheroids have positive $A_{\rm IA}$, indicating an alignment between the major axis of their projected shape and tidal field. Low mass disks have an $A_{\rm IA}$ that is consistent with zero;
    low mass spheroids have a weak, redshift-dependent shape alignment. In contrast, we found a `spin flip' signal in TNG300 at two-point statistics level. The sign of $P_{\delta E_J}$ for the low-mass disks and spheroids ($z=0.3, 0.5$) is reversed at $k<1h{\rm Mpc^{-1}}$ relative to that of the massive disks and spheroids, indicating a flip of the angular momentum vector of the low mass galaxies. The spin of low-mass disks and spheroids ($z=0.3, 0.5$) is parallel with the filament, while massive spheroid and disks show a clear spin alignment signal that is perpendicular to the filament. The morphology dependent IA signal can be tested using future observations.
    
    \item The IA also varies for central and satellites. Centrals have their shape and spin aligned with the tidal field at both large and small scales. The small scale alignment reflects that centrals aligned with their host halos. The good agreement between $P_{\delta E}$ and $P_{\delta E_J}$ for massive centrals also indicates a strong alignment between their spin and shape; while low mass centrals have no detection of spin alignment indicated by $P_{\delta E_J}$. Low-mass satellites, on the other hand, show an inverted shape and spin alignment at large scales. The small scale IA of satellites is mainly driven by the radial alignment of satellites in their host halo. The various predictions on central/satellite IA by different models will enable us to constrain the related physics when future observations such as Subaru PFS are available.

    \item We made a prediction for the 3D IA power spectrum measurement by combing the ongoing/forthcoming image and spectroscopic surveys. The IA power spectrum between the ellipticity field traced by massive red galaxies, i.e., LRGs, and the density field traced by either blue galaxies or star-forming galaxies can be measured with a promising S/N ratio. The S/N ratio can be improved significantly if the number density of the density field tracer is larger, highlighting an important role of deep survey in IA observations.

\end{itemize}
One caveat of our study is that the IA alignment of galaxies relies on the galaxy formation model employed in the simulation, thus the prediction made by this work might not be valid for other galaxy formation models. This model dependence, on the other hand, provides another avenue to constrain the physical models used in varying works. It would be worth measuring the IA power spectrum for different types of galaxies from actual data and using it to constrain or test the galaxy physics employed in the TNG simulations. This would be our future work.

\acknowledgments
We thank the IllustrisTNG team for making their simulation data publicly available. The IllustrisTNG simulations were undertaken with compute time awarded by the Gauss Centre for Supercomputing (GCS) under GCS Large-Scale Projects GCS-ILLU and GCS-DWAR on the GCS share of the supercomputer Hazel Hen at the High Performance Computing Center Stuttgart (HLRS), as well as on the machines of the Max Planck Computing and Data Facility (MPCDF) in Garching, Germany. We thank Ananth~Tenneti for helpful discussion.
This work was supported in part by World Premier International
Research Center Initiative (WPI Initiative), MEXT, Japan, and JSPS
KAKENHI Grant Numbers JP15H05887, JP15H05893, JP15K21733, JP17K14273, JP19H00677, 
and JP20J22055, and by JST AIP Acceleration Research Grant Number JP20317829, Japan. T.K. is supported by JSPS Research Fellowship for Young Scientists.
KO is supported by JSPS Overseas Research Fellowships.
Y.K. is supported by the Advanced Leading Graduate Course for Photon Science at the University of Tokyo.

\appendix
\counterwithin{figure}{section}

\section{Selection of $M_\star$-limited and SFR-limited Samples}
\label{sec:select_fixng}

\begin{figure}[!htb]
\centering 
\includegraphics[width=.45\textwidth]{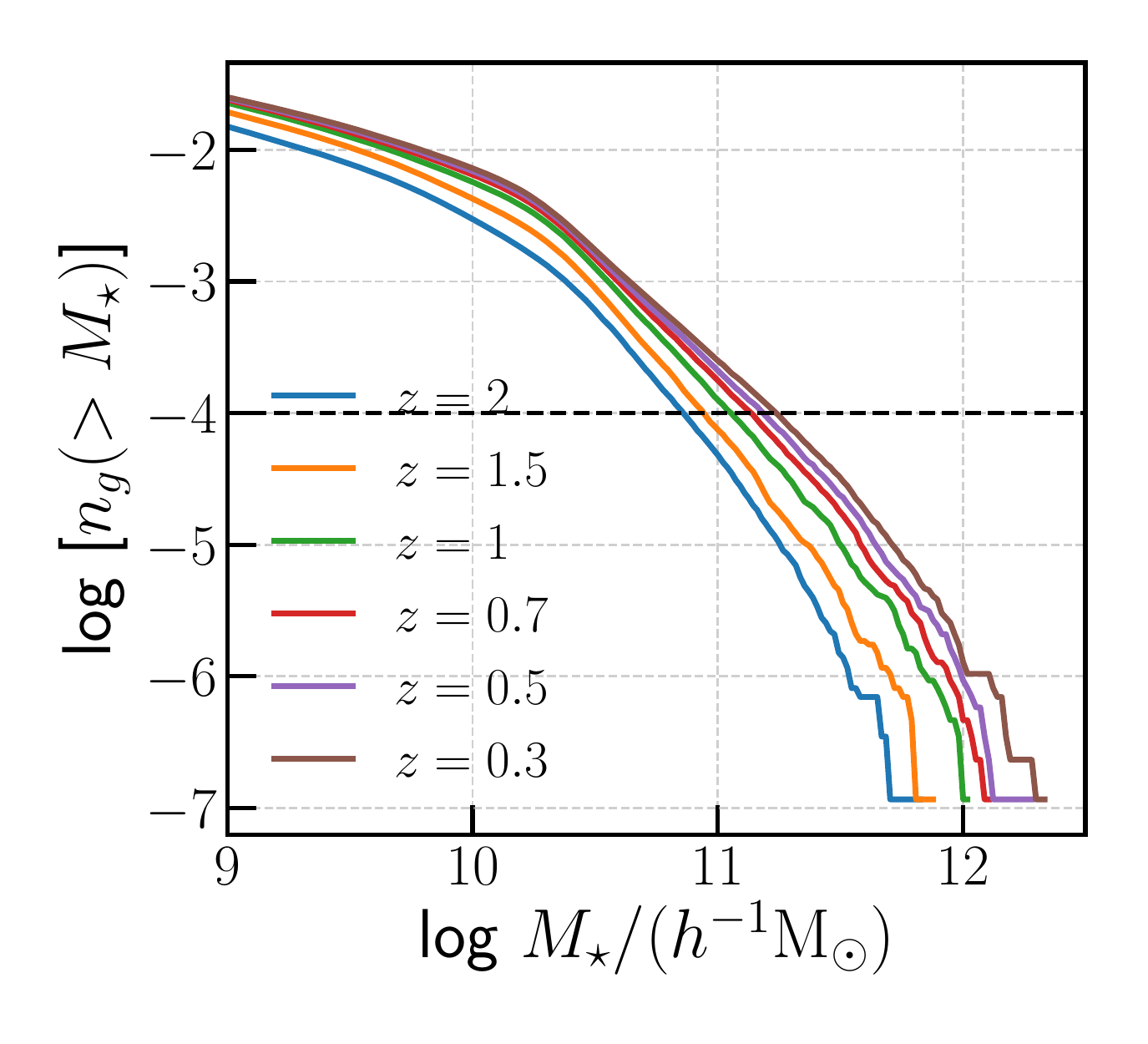}
\includegraphics[width=.45\textwidth]{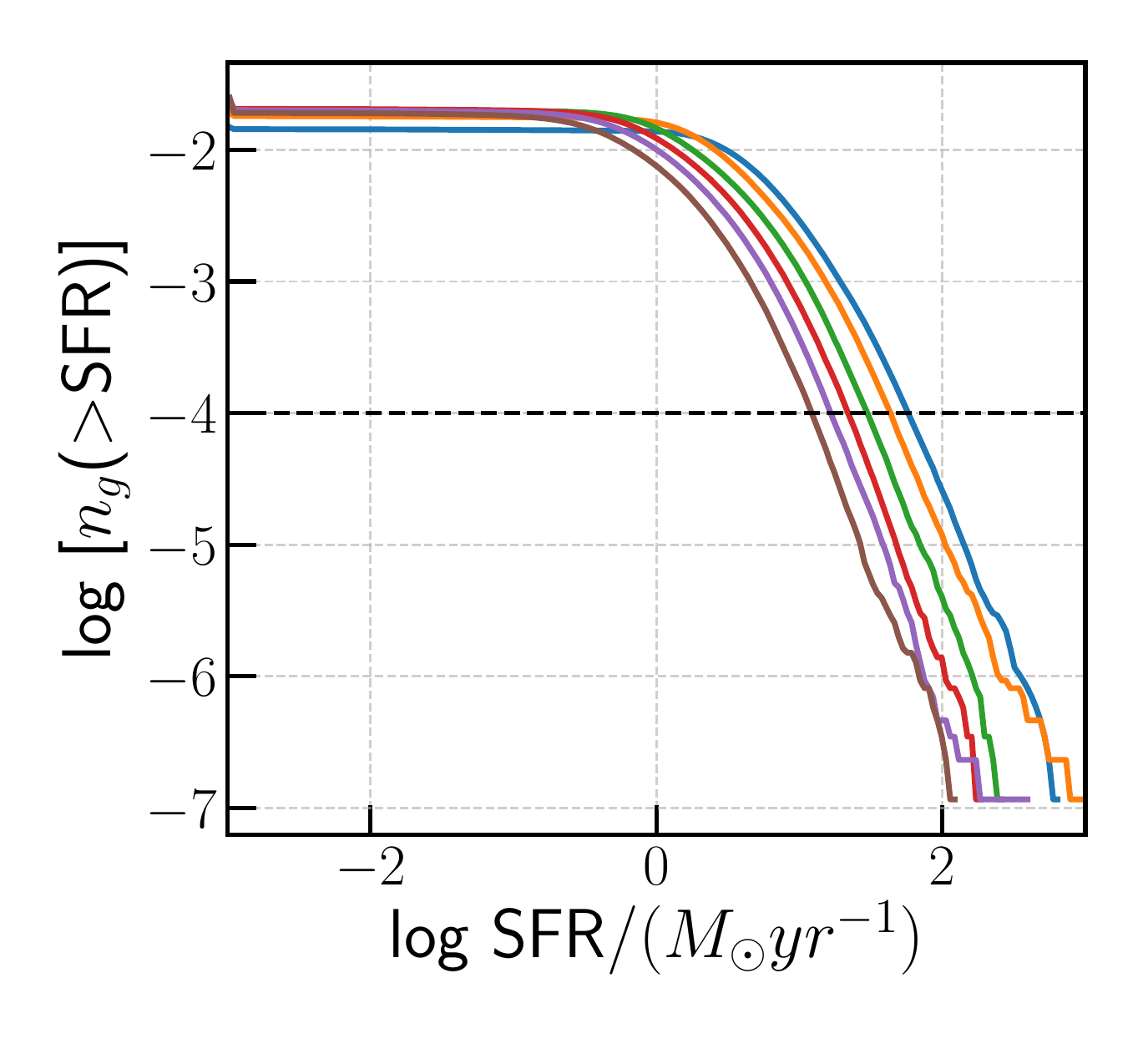}
\caption{\label{fig:ng_fixed} Selection of the $M_\star$-limited and SFR-limited samples. The lines of varying colors show the cumulative number density distributions of galaxies in TNG300 from redshift $0.3$ to $2$, ranked by the stellar mass (left panel) and SFR (right panel) separately. The dashed lines indicate the number density threshold of $10^{-4}(h^{-1}{\rm Mpc})^{-3}$.}
\end{figure}

We present a brief illustration of our methods for selecting the $M_\star$-limited and SFR-limited samples. We ranked the total galaxies in our sample by $M_\star$ or SFR. Here the SFR is the summing up of the SFR of all gas cells within the galaxy. Fig.~\ref{fig:ng_fixed} shows the cumulative number density distribution at each redshift ranked either by $M_\star$ or SFR. Then we decide the thresholds above which the number density of galaxies reaches $n_g=10^{-4}(\mpc)^{-3}$. As the redshift decreases, the stellar mass of galaxies grows and more galaxies are quenched. Consequently, the $M_\star$ threshold that corresponds to $n_g(>M_\star)=10^{-4}(\mpc)^{-3}$ becomes larger, while the SFR threshold becomes smaller.

\section{Effects of Different Inertia Tensor Measurement}
\label{sec:varying_e_def}

\begin{figure}[!htb]
\centering 
\includegraphics[width=.8\textwidth]{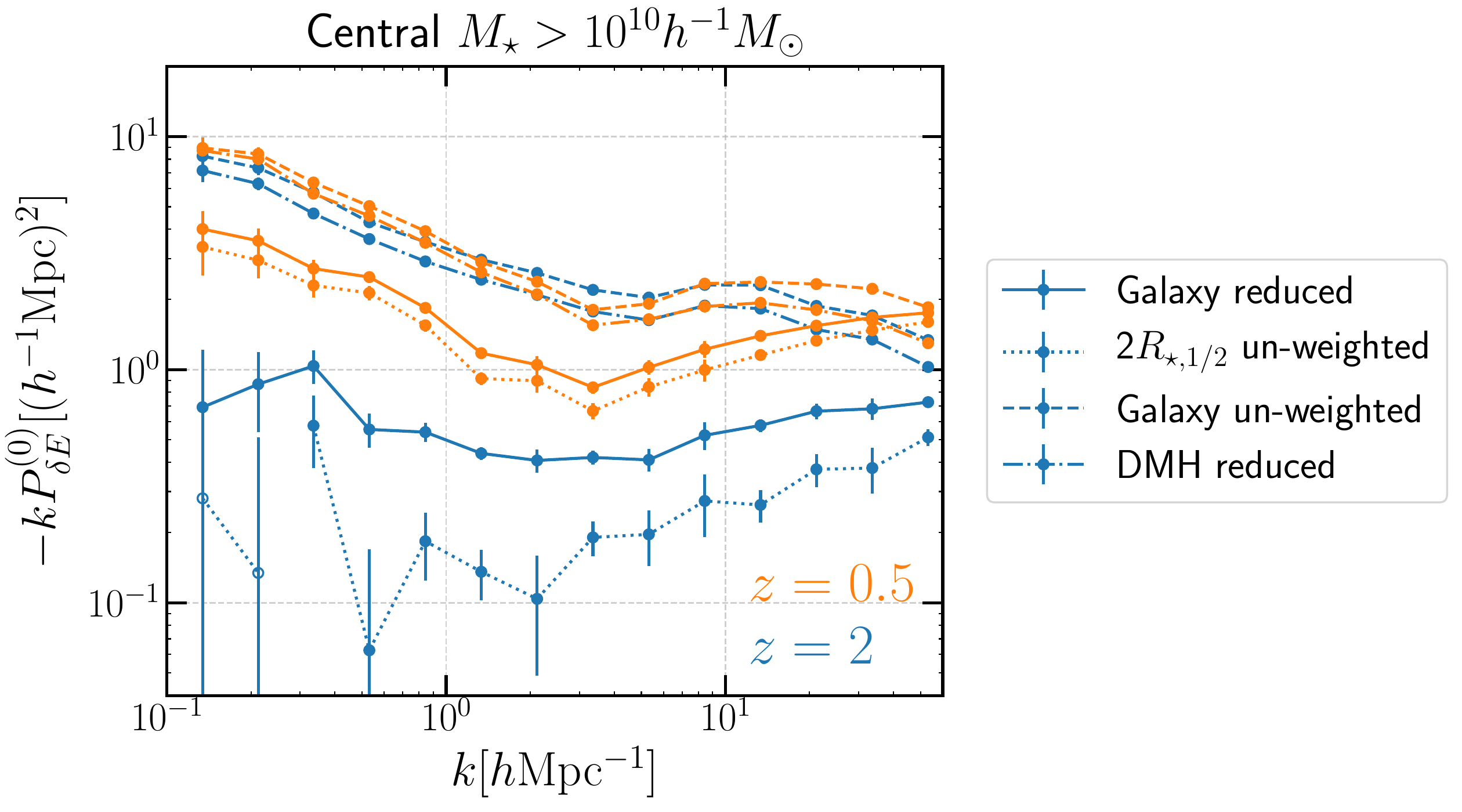}
\caption{\label{fig:PS_diffE} The IA power spectrum of $M_\star>10^{10}\msun$ central galaxies measured using different inertia tensor definitions: using all the star particles within the galaxy and weighted by $1/r^2$ - `Galaxy reduced'; using only star particles within $2R_{\star,1/2}$, without any extra weighting - `2$R_{\star,1/2}$ un-weighted'; using all the star particles within the galaxy without any extra weighting - `Galaxy un-weighted'; using the dark matter particles within the subhalo of the galaxy and weighted by $1/r^2$-`DMH reduced'. The filled symbols are for $-kP^{(0)}_{\delta E}$, while the open symbols are for $kP^{(0)}_{\delta E}$.}
\end{figure}

The IA power spectrum measured in the simulation depends on how exactly the inertia tensor is measured for individual galaxy. In this section, we compare the effects of different measurement methods on the IA power spectrum. Besides the one used in the main text, i.e., the reduced inertia tensor calculated by using all the stellar particles within the galaxy (`Galaxy reduced'), we can also measure the inertia tensor without the $1/r^2$ weighting (`Galaxy un-weighted'), i.e.,
\begin{equation}
    I_{ij}=\frac{\sum_{n} m_n x_{ni} x_{nj}}{\sum_{n} m_n},  
\end{equation}
or using only the stellar particles within $2R_{\star, 1/2}$ without the $1/r^2$ weighting (`Within $2R_{\star, 1/2}$ un-weighted'), or using the dark matter particles within the subhalo weighted by $1/r^2$ (`DMH reduced').
The `un-weighted' ellipticity up-weights contributions from outer (stellar or/and dark matter) particles in each system.

Fig.~\ref{fig:PS_diffE} shows the IA power spectrum of galaxies ($M_\star>10^{10}\msun$) measured using different inertia tensor definitions. Note that we use the same sample of galaxies, and the differences in the IA power spectrum are from the different ellipticity definitions. The resulting IA signal for the `Galaxy un-weighted' ellipticity method is significantly higher than those for other methods over all the scales we consider. We should note that the $k$ dependence of IA signals for $z=0.5$ on large scales ($k\lesssim 1~\hmpci$) is similar for all the methods, supporting that the large-scale IA signal is mainly captured by the 
constant $A_{\rm IA}$ coefficient. In other words, the different ellipticity method changes mainly the amplitude of $A_{\rm IA}$ for the large-scale IA signal \citep[also see Ref.][for the smilar discussion]{2020arXiv200412579K}. However, the difference can not be fully explained by the difference in the ellipticities. 
For example, the RMS of intrinsic ellipticities measured by `Galaxy reduced', `Galaxy un-weighted', `Within $2R_{\star, 1/2}$ un-weighted', and `DMH reduced' for the sample are: $e_{\rm rms}=0.30, 0.34, 0.34, 0.25$ at $z=2$, and $e_{\rm rms}=0.29, 0.29, 0.31, 0.20$ at $z=0.5$, respectively.
Thus the rms ellipticities are not so largely changed for the different methods. These mean that the IA power spectrum can be measured with higher signal-to-noise ratios, if we can reliably use the ellipticity observables upweighting the outer region of individual galaxies.
However, in practice, this is not obvious to easily achieve in actual observations, because the outer regions of galaxies are more affected by systematic effects in photometry such as sky subtraction or flat fielding. 
The amplitude and the shape of the IA power spectrum measured using `DMH reduced' generally follow that of `Galaxy un-weighted', providing a strong support for the methodology of using the inner part of the dark matter halo to represent the shape of galaxies in the $N$-body simulation\cite{2020arXiv200412579K}. However we should keep in mind that a misalignment between shapes of galaxies and dark matter halos should exist to some degree, and the degree of misalignment could vary with different radii inside the halo. 
This might explain the slight difference ($\sim0.1$dex) between the IA power spectrum of those two methods.

In addition, the effects of different ellipticity measurement is redshift dependent. At $z=0.5$, the IA power spectrum of the `Galaxy reduced' is close to that of the `Within $2R_{\star, 1/2}$ un-weighted', which is reasonable since both of them weight the inner region more. In comparison, at $z=2$, the IA power spectrum of those two are really different from each other. At large scales, when `Within $2R_{\star, 1/2}$' is adopted, the IA signal is even reversed. This indicates that the stellar mass distribution (or morphology) of the galaxies at $z=2$ could be really different even when different inner regions are considered. During this epoch, galaxies would be still in the rapid evolution stage consisting of building blocks in the outer regions, and the morphology is not yet established. Thus for such a high redshift, if we can define the galaxy ellipticity including building blocks in the outer region in a reliable manner, we could achieve a significant detection of the IA power spectrum. This is an interesting possibility, and worth exploring.

\section{Intrinsic Alignment Strength $b_K$}
\label{sec:b_K}

In analogy with the expansion of galaxy overdensity field in terms of the underlying matter field with bias coefficients, the linear alignment model can be rephrased in terms of the tidal field, which has the same dimension as the density fluctuation field, as
\begin{align}
g_{ij}(\bm{x},z)=b_K K_{ij},
\end{align}
where we defined the effective tidal field as
\begin{align}
K_{ij}(\bm{x},z)\equiv \left(\nabla^{-2}\partial_i\partial_j-\frac{\delta^K_{ij}}{3}\right)\delta(\bm{x},z).
\end{align}
Here we follow the notations in Ref.~\cite{Schmidtetal2015}; $g_{ij}$ is the three-dimensional shear field at a galaxy position, and $b_K$ is a dimension-less quantity. In this expansion, we can consider $b_K$ as the linear IA coefficient relating the IA shear of galaxy shapes to the underlying matter field, very much like the linear density bias parameter given by $\delta_g=b_1\delta$ on large scales \citep{Schmidtetal2015}. 
However, an actual observable for each galaxy is the projected shape shear of an each galaxy image on the sky (see Section~\ref{subsec:e_measurement}). The shape shear for each galaxy is expressed from $(2\times 2)$ submatrix of the 3D shear matrix as
\begin{align}
\gamma^I_{(+,\times)}\equiv \left(\frac{g_{11}-g_{22}}{2},g_{12}\right)=b_K\left(\frac{K_{11}-K_{22}}{2},K_{12}\right).
\end{align}
Using these redefined quantities, the IA power spectrum can be written as
\begin{align}
&P_{\delta E}(k,\mu;z)=\frac{b_K}{2}(1-\mu^2)P^{\rm lin}_\delta(k,z), \nonumber\\
&P_{EE}(k,\mu;z)=\frac{b_K^2}{4}(1-\mu^2)^2P^{\rm lin}_\delta(k,z), \nonumber\\
&P_{{\rm g}E}(k,\mu;z)=\frac{b_K}{2}b_{\rm g}(1-\mu^2)P^{\rm lin}_\delta(k,z).
\end{align}

\begin{figure}[!htb]
\centering 
\includegraphics[width=.9\textwidth]{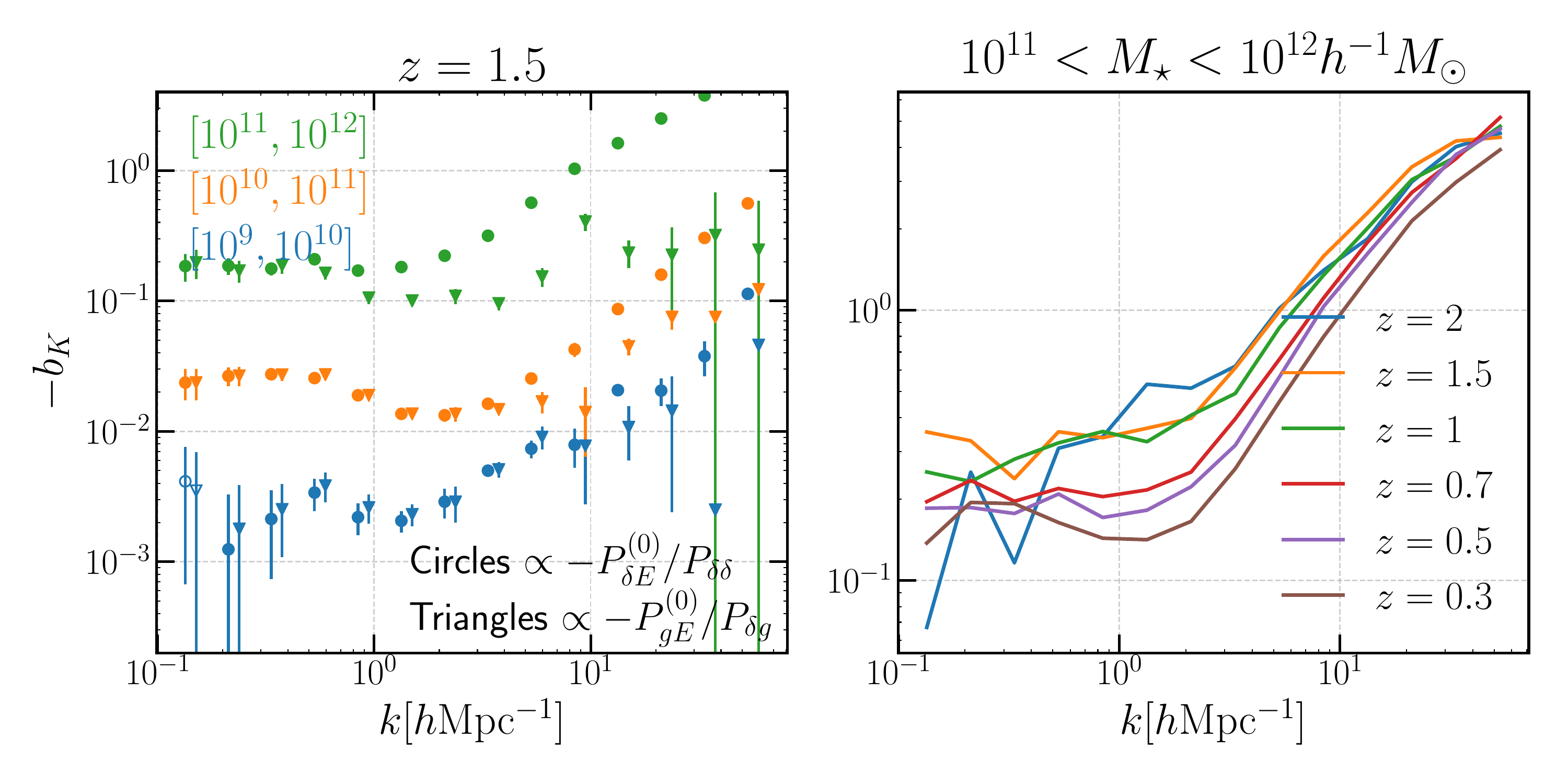}
\caption{\label{fig:B_E0_mstarz} Left: $b_K\propto P_{\delta E}^{(0)}/P_{\delta\delta}$ (circles) and $b_K\propto P_{g E}^{(0)}/P_{\delta g}$ (triangles) for galaxies in three different stellar mass ranges at $z=1.5$. Right: $b_K$ for the galaxies in the mass range of $10^{11}<M_\star<10^{12}\msun$ from $z=0.3$ to $z=2$.}
\end{figure}

Fig.~\ref{fig:B_E0_mstarz} presents $b_K$ for three stellar mass bins at $z=1.5$ (left panel) and the redshift evolution of $b_K$ for massive galaxies with $10^{11}<M_\star/\msun<10^{12}$. 
Thus, $b_K$ has the same stellar mass dependence as $A_{\rm IA}$ at given redshift. Besides, $b_K$ has a redshift dependence driven by the growth of matter clustering; $|b_K|$ is higher at higher redshifts for a galaxy sample of a fixed stellar mass.

\bibliographystyle{JHEP}
\bibliography{refs}
\end{document}